\documentclass[twocolumn,showpacs,preprintnumbers,amsmath,amssymb]{revtex4}
\usepackage{epsfig}
\usepackage{dcolumn}
\usepackage{bm}

\newcommand{\remove}[1]{}

\def\be{\begin{equation}}
\def\ee{\end{equation}}

\newcommand{\beq}{\begin{equation}}
\newcommand{\eeq}{\end{equation}}
\newcommand{\beqa}{\begin{eqnarray}}
\newcommand{\eeqa}{\end{eqnarray}}

\newcommand{\ii}{{\rm i}}

\newcommand{\vx}{{\bf x}}
\newcommand{\vk}{{\bf k}}

\renewcommand{\vr}{{\bf r}}

\newcommand{\bea}{\begin{array}}
\newcommand{\ea}{\end{array}}

\begin{document}

\title{Non-screening of the Cosmological Background in K-mouflage modified gravity}

\author{Philippe Brax}
\affiliation{Institut de Physique Th\'eorique, Universit\'e Paris-Saclay, CEA, CNRS, F-91191 Gif-sur-Yvette Cedex, France}
\author{Patrick Valageas}
\affiliation{
Institut de Physique Th\'eorique, Universit\'e Paris-Saclay, CEA, CNRS, F-91191 Gif-sur-Yvette Cedex, France}
\vspace{.2 cm}

\date{\today}
\vspace{.2 cm}

\begin{abstract}

We describe the effects of the cosmological background on the K-mouflage screening
properties of an astrophysical structure.
We show that the K-mouflage screening of the spatial gradients of the scalar field,
i.e. the screening of the fifth force, happens inside a dynamically generated screening
radius. This radius is smaller than the location where the quasistatic approximation, i.e.
where the spatial gradients exceed the time derivative, holds. Even though
this quasistatic radius is much smaller than the size of the matter overdensity,
spatial gradients remain well described by the quasistatic approximation up to the horizon.
However, cosmologically we find that the time derivatives can remain dominant at redshifts
$z\gtrsim 2$, when the cosmic web shows a faster growth.
Despite the existence of K-mouflage screening, we confirm that the values of the scalar field itself
are still dominated by the cosmological background, down to the center of the matter
overdensity, and that for instance the time drift of Newton's constant due to
the large-scale cosmological evolution highly constrains K-mouflage models.

\end{abstract}

 \vskip2pc

\maketitle

\section{Introduction}
\label{sec:Introduction}

Scalar models with derivative actions and a coupling to matter, such as K-mouflage \cite{Babichev:2009ee,Brax:2012jr,Brax:2014c,Brax:2014wla} and Galileon-like theories \cite{Nicolis:2008in,Vainshtein:1972sx}, screen fifth force effects in the presence of matter.
This is due to the nonlinearities in the kinetic terms of the scalar field.
This is sufficient to guarantee that most Solar System tests of gravity are fulfilled by these
models.
Now that the observation of the equality, up to a very high accuracy, between the speeds of
gravity and light has ruled out most Horndeski models with self-tuning properties
\cite{Creminelli:2017sry},
K-mouflage remains a serious alternative to the $\Lambda$-CDM paradigm.
Of course, K-mouflage models do not propose a solution to the ``old'' cosmological constant
problem \cite{Weinberg:1988cp}, but their peculiar features on the growth of structures are
sufficiently compelling to motivate further studies, in particular on the influence
on the large-scale cosmological evolution and its backreaction on small-scale
properties \cite{Barreira2015}.
This is the case of the time drift of Newton's constant, due to the absence of screening
by the K-mouflage mechanism of the time dependence of the scalar field.
In this paper, we characterize this property by going beyond the usual
quasistatic approximation, which assumes that any slow dependence on time
of the background scalar field can be added to the static profile associated with dense
objects. We analyze the nonlinear regime with a fully time-dependent cosmological solution
describing the matter era. We show how, when screening of the spatial gradients occurs
inside an overdensity, the time drift itself is not affected.

In section~\ref{sec:models}, we define the K-mouflage models by the nonlinear
Klein-Gordon equation that governs the evolution of the scalar field.
In section~\ref{sec:linear}, we consider the situation with nonscreening, which
corresponds to a standard kinetic term, and we study how the cosmological background
propagates down to the center of the overdensity while spatial gradients converge to the
quasistatic limit on subhorizon scales.
In section~\ref{sec:nonlinear}, we investigate how the situation is modified
by the screening effects due to the nonlinearities associated with large field gradients.
In section~\ref{sec:Conclusion} we conclude.

\section{K-mouflage models}
\label{sec:models}
\subsection{The dynamics}

The scalar field $\phi$ in K-mouflage models  obeys the nonlinear Klein-Gordon equation
\cite{Brax:2014a}
\beq
\nabla_{\mu} \left[ K' \nabla^{\mu} \phi \right] = \frac{\beta\rho}{M_{\rm Pl}} ,
\label{eq:KG-1}
\eeq
where $\nabla_{\mu}$ is the covariant derivative with respect to
the Einstein-frame metric $g_{\mu\nu}$, $\rho$ the matter density
and $\beta$ the coupling constant. The function $K$ is a function of the kinetic term
$\chi=-(\partial \phi)^2/(2 {\cal M}^4)$, where ${\cal M}^4$ is of the order
of the dark-energy scale.

For the cosmological background, or on large cosmological scales,
matter density fields and the scalar field are exactly or almost homogeneous, so that
$\chi$ is dominated by the time derivative and $\chi>0$.
In the vicinity of static compact objects, such as stars, or in high-density regions
such as the cores of galaxies, spatial gradients dominate over time derivatives
and $\chi<0$.
Thus, the high-density cosmological background associated with the early Universe
corresponds to $\chi \to +\infty$, whereas the high-density regions associated with
quasistatic astrophysical objects correspond to $\chi \to -\infty$.
This corresponds to  two unrelated screening regimes, if the function $K$ is nonlinear
for both large positive and negative argument.

The faster-than-linear growth of $K$ for $\chi \to +\infty$, i.e. $K' \to +\infty$, ensures that
the scalar-field energy density is negligible at high redshift as compared with the
matter density, so that one recovers the standard cosmology \cite{Brax:2014a}.
For small values of $\chi$, associated with low redshifts, we expand
$K=-1+\chi+...$ (the unit factors define the normalization of ${\cal M}^4$ and
$\phi$) and we recover a canonically normalized scalar field (the linear term)
with a cosmological constant (the constant term $-1$).

In a quasistatic high-density region, or close to a compact astrophysical object,
spatial gradients become large and a screening mechanism also comes into play
if $K'$ becomes large for large negative $\chi$
\cite{Babichev:2009ee,Brax:2014c}.
This slows down the growth of the scalar field gradients with the rise of the
matter density.
For instance, in a static spherically symmetric overdensity,
Eq.(\ref{eq:KG-1}) gives after one integration an equation of the form
$K' d\phi/dx \propto M(<x)/x^2$, where $M(<x)$ is the mass inside the radius $x$,
so that the scalar field gradient
is suppressed by a factor $1/K'$. This gives rise to the K-mouflage screening
mechanism and allows the fifth force to become negligible as compared with
the Newtonian gravity in small and high-density regions.

If we assume that such a local picture fully describes the behavior of the scalar field
in small-scale high-density regions, we could expect that in a similar fashion
the large value of $K'$ should suppress all derivatives of $\phi$, the time derivative
as well as spatial derivatives. This is for instance the behavior that is obtained
by multiplying $K'$ in Eq.(\ref{eq:KG-1}) by a large constant factor.
Then, the scalar field at the center of a high matter overdensity should decouple
from the cosmological background and no longer evolve
inside a static matter halo.
It turns out that this picture is not correct.

In this paper, we investigate in more details this issue, using simple power-law
density profiles, for which we can derive explicit analytical results. We find that although
spatial gradients are well predicted by the quasistatic approximation
on subhorizon scales, the scalar field itself does not truly decouple from
the cosmological background.
Its time derivative remains greater than the spatial gradients down to scales
much below the size of the matter overdensity, and its value at the center
closely follows the drift of the cosmological background.

\subsection{Rescaled variables}
\label{sec:rescaled}

Neglecting the metric fluctuations from the Friedmann-Lema\^itre-Robertson-Walker
(FLRW) background, with scale factor $a$, the nonlinear Klein-Gordon
equation (\ref{eq:KG-1}) reads
\beq
- a^{-4} \partial_\tau ( a^2 K' \partial_\tau\phi ) + a^{-2} \nabla ( K' \nabla \phi)
=  \frac{\beta \rho}{M_{\rm Pl}} ,
\label{eq:KG-2}
\eeq
where $\tau$ is the conformal time and $\nabla=\partial_{\bf x}$
the gradient with respect to the comoving coordinate ${\bf x}$.
For simplicity, we consider an Einstein-de Sitter universe, i.e. matter dominated, with
\beq
a= \left(\frac{t}{t_0}\right)^{2/3} = \left(\frac{\tau}{\tau_0}\right)^2 ,
\;\;\; t_0 = \frac{2}{3 H_0} , \;\;\; \tau_0 = \frac{2}{H_0} ,
\;\;\; \bar\rho = \frac{\bar\rho_0}{a^3} ,
\label{eq:EdS}
\eeq
where $t_0$ is the age of the universe at redshift $z=0$
and $\tau_0$ the conformal time today.
It is convenient to introduce the dimensionless coordinates
\beq
\tilde\tau = \frac{\tau}{\tau_0} , \;\;\;
\tilde{\bf x} = \frac{\bf x}{\tau_0} , \;\;\;
\tilde\phi = \frac{\phi}{M_{\rm Pl}} .
\label{eq:rescaled}
\eeq
Then, the Klein-Gordon equation (\ref{eq:KG-2}) reads
\beq
- \partial_{\tilde\tau} ( K' \partial_{\tilde\tau}\tilde\phi )
- \frac{4}{\tilde\tau} K'\partial_{\tilde\tau}\tilde\phi
+ \tilde\nabla ( K' \tilde\nabla \tilde\phi)
= 12 \beta \frac{\rho}{\bar\rho \tilde\tau^2}
\label{eq:KG-3}
\eeq
and the argument of the kinetic function $K$ is
\beq
\chi = \frac{1}{2\tilde\tau^4} \left[ (\partial_{\tilde\tau}\tilde\phi)^2
- (\tilde\nabla\tilde\phi)^2 \right] ,
\label{eq:chi-tilde}
\eeq
with the choice of normalization ${\cal M}^4 = M_{\rm Pl}^2 H_0^2/4$.
In the following we will omit the tildes and only work with these rescaled
quantities.
In this paper we focus on the response of the scalar field to the cosmological
background and matter overdensities. Therefore, we discard the backreaction
of the scalar field onto the cosmological expansion history and the formation of
matter overdensities.
This also corresponds to a small coupling constant $\beta \ll 1$.
This is actually the case of interest as  observations show that the
fifth force must remain subdominant as compared with Newtonian gravity
and we must recover the standard cosmological expansion up to an accuracy
of a few percents at low redshifts.
We will study the evolution of the scalar field for a given cosmological background,
defined by the Einstein-de Sitter solution (\ref{eq:EdS}), and for given matter
overdensities.

\subsection{Physical radial coordinate}
\label{sec:physical-radius}

On small astrophysical scales or in the laboratory, where we usually neglect
the expansion of the Universe, we use the physical coordinate ${\bf r}= a {\bf x}=
\tau^2 {\bf x}$.
For spherical profiles, the Klein-Gordon equation (\ref{eq:KG-3}) becomes
in the coordinates $\{r,\tau\}$,
\beqa
&& \frac{1}{r^2} \partial_r \left[ r^2 K' \partial_r \phi \right]
 -\frac{1}{\tau^4} \partial_\tau \left[ K' \left( \frac{2 r}{\tau} \partial_r \phi
+ \partial_\tau \phi \right) \right] \nonumber \\
&&  - \frac{2}{r \tau^5} \partial_r \left[ r^2 K' \left( \frac{2 r}{\tau} \partial_r \phi
+ \partial_\tau \phi \right) \right] = 12 \beta \frac{\rho}{\bar\rho_0} ,
\label{eq:KG-r}
\eeqa
while the kinetic argument $\chi$ reads as
\beq
\chi = \frac{1}{2} \left[ \frac{1}{\tau^4} ( \partial_\tau \phi )^2
+ \frac{4 r}{\tau^5} \partial_\tau \phi \partial_r \phi
+ \left( \frac{4 r^2}{\tau^6} -1 \right) ( \partial_r \phi )^2 \right] .
\label{eq:chi-r}
\eeq
We recover the Klein-Gordon equation of Minkowski spacetime,
$-\partial_t(K' \partial_t \phi) + r^{-2} \partial_r( r^2 K' \partial_r \phi ) =
12 \beta \rho / \bar\rho_0$, on small subhorizon scales $r \ll \tau^3$,
for small time scales $\Delta\tau \ll \tau$.
Because of the expansion of the Universe, which gives the relation
$\left. \partial_\tau \phi \right|_{\bf x} = \frac{2 r}{\tau} \partial_r \phi
+ \left. \partial_\tau \phi \right|_{\bf r}$, Eq.(\ref{eq:KG-r}) displays
a mixing of spatial and time derivatives, even when $K'$ is a constant.
Then, although we consider in this paper the relaxation of the scalar field
around a cosmological matter overdensity that virializes to a static
profile on small scales, it usually remains more convenient to work with the
comoving Klein-Gordon equation (\ref{eq:KG-3}).

\section{Standard kinetic term}
\label{sec:linear}

\subsection{Cosmological background}
\label{sec:background}

For the homogeneous cosmological background, Eq.(\ref{eq:KG-3}) can be integrated
once to give
\beq
\bar{K}' \frac{d\bar\phi}{d\tau} = - \frac{4\beta}{\tau} .
\label{eq:KG-background}
\eeq
In this paper, we are not interested in the screening of the cosmological background
at high redshifts. Therefore, we can take $\bar K'$ to be constant
for the cosmological background and choose the normalization $\bar K'=1$.
This corresponds to kinetic functions with $K'=1$ for $\chi \geq 0$,
or to the standard kinetic term $K(\chi)=\chi$.
This gives the cosmological background solution
\beq
\bar{K}'=1 : \;\;\; \frac{d\bar\phi}{d\tau} = - \frac{4\beta}{\tau} , \;\;\;
\bar\phi = - 4 \beta \ln\tau .
\label{eq:phi-bar}
\eeq

Such models do not produce a self-acceleration of the Universe that is
significantly different from a cosmological constant. Indeed, the acceleration
arises from the nonzero negative value of $K(\chi)$ at $\chi=0$,
which we can set equal to $-1$ while the scale ${\cal M}^4$ in the Lagrangian
${\cal L}_\phi = {\cal M}^4 K(\chi)$ is set to the observed dark energy scale.
However, at this level this is a matter of definition, and one can as well set
$K(0)$ to zero and interpret ${\cal M}^4$ as a standard cosmological constant.

It is possible to obtain a slightly more genuine self-acceleration with models such that
$K'(\chi)$ vanishes for a value $\chi_\star > 0$ \cite{Brax:2014a}.
During the cosmological evolution $\chi$ decreases towards $\chi_\star$,
which is only reached in the infinite future, and the self-acceleration is provided by the nonzero
value $K(\chi_\star)<0$ at this fixed point.
However, one could again interpret $K(\chi_\star)$ as a standard cosmological constant.
Models with $K'<0$ show strong ghost instabilities, which would imply a very low
cutoff for the theory (typically below 1 keV) \cite{Brax:2014a}, therefore one requires
$K'>0$ for $\chi>\chi_\star$.
However, the range $\chi \lesssim \chi_\star$ where $K'$ could become negative could
remain problematic.
We do not discuss further these models and the cosmological evolution here.
Indeed, we are not interested in the cosmological evolution itself, but only in the
impact of its time dependence on the small-scale regime, which corresponds
to the different range $\chi < 0$.

\subsection{General linear solution}

In this section, we consider the case of the standard kinetic function,
where $K'=1$ for all positive and negative $\chi$. Then, the Klein-Gordon equation
(\ref{eq:KG-3}) is linear and reads as
\beq
- \partial^2_\tau \phi_L - \frac{4}{\tau} \partial_\tau \phi_L + \nabla^2 \phi_L
 = 12 \beta \frac{\rho}{\bar\rho \tau^2} .
\label{eq:KG-linear}
\eeq
To distinguish from the nonlinear case studied in section~\ref{sec:nonlinear} below
for varying $K'$, we added the subscript ``L''.
This recalls that for constant $K'$ the Klein-Gordon equation
is linear.
Note that this does not involve any perturbative expansion over the density contrast
or the scalar field (we only neglect the fluctuations of the FLRW metric
and consider the matter density as an external source).
To work with functions that vanish at infinity, we subtract the cosmological
background by defining
\beq
\phi_L = \bar\phi + \varphi_L , \;\;\; \rho = \bar\rho (1+\delta) ,
\label{eq:background-split}
\eeq
where $\varphi_L$ and $\delta$ are not necessarily small but vanish at large distances.
Indeed, in this paper we are interested in the formation of nonlinear structures,
with a finite size, amidst  the cosmological background.
Then, the deviation $\varphi_L$ obeys the linear equation
\beq
{\cal O} \cdot \varphi_L = 12 \beta \frac{\delta}{\tau^2} ,
\label{eq:KG-linear-pert}
\eeq
where we have introduced the linear operator ${\cal O}$ defined by
\beq
{\cal O} =  - \partial^2_\tau - \frac{4}{\tau} \partial_\tau + \nabla^2 .
\label{eq:O-def}
\eeq
Using the associated retarded Green function
\beq
{\cal O} \cdot {\cal G}(\vx,\tau;\vx',\tau') = \delta_D(\vx-\vx') \delta_D(\tau-\tau') ,
\label{eq:Green-def}
\eeq
we can solve the linear equation (\ref{eq:KG-linear-pert}) as
\beq
\varphi_L(\vx,\tau) = 12 \beta \int d\vx' d\tau' {\cal G}(\vx,\tau;\vx',\tau')
\frac{\delta(\vx',\tau')}{\tau'^2} .
\label{eq:phi-linear}
\eeq
Solving Eq.(\ref{eq:Green-def}) by using its Fourier transform, we obtain
\beqa
{\cal G}(\vx,\tau;\vx',\tau') & = & \theta(\tau-\tau') \int \frac{d\vk}{(2\pi)^3}
e^{\ii \vk\cdot(\vx-\vx')} \frac{k\tau'^3}{\tau} \nonumber \\
&& \times [ n_1(k\tau') j_1(k\tau) - j_1(k\tau') n_1(k\tau) ] , \hspace{0.7cm}
\label{eq:Green-j1-n1}
\eeqa
where $\theta$ is the Heaviside function, $j_1$ and $n_1$ are the spherical
Bessel functions of the first and second kind.
Substituting the explicit expressions of $j_1$ and $n_1$ in terms of cosines
and sines, we can easily check that in the limit of small lengths and timescales,
$|\vx-\vx'| \to 0$, $\tau-\tau'\to 0$, $k \to \infty$, we recover the usual Green function
of the 3D wave equation \cite{Morse:1953aa},
\beq
{\cal G} \to - \frac{\theta(\tau-\tau') \delta_D(|\vx-\vx'|-(\tau-\tau'))}{4\pi |\vx-\vx'|} .
\label{eq:Green-standard}
\eeq
This corresponds to the limit where the Hubble friction term in the operator (\ref{eq:O-def})
is negligible.

\subsection{Self-similar matter density profiles}
\label{sec:self-similar}

We now investigate how the scalar field reacts to the formation of an overdense region.
We consider a class of  simple cases where we can obtain explicit expressions,
the self-similar spherical power-law density profiles
\beq
\delta(\vx,\tau) = \left( \frac{x}{x_s(\tau)} \right)^{-\gamma}, \;\;\;
x_s(\tau) = x_\star \tau^{\alpha} , \;\;\; x_\star \ll 1 .
\label{eq:delta-self-similar}
\eeq
In the rescaled coordinates (\ref{eq:rescaled}), the time $\tau$ runs over
$0\leq\tau\leq 1$, and the condition $x_\star \ll 1$ ensures that the overdensity always
remains far inside the Hubble radius.

The profile (\ref{eq:delta-self-similar}) corresponds to a halo of inner density slope
$\gamma$ and size $x_s(\tau)$, which grows with time in a self-similar fashion.
Such a solution can be achieved for instance by the collapse of a polytropic
gas with a power-law initial linear density contrast profile
\cite{Teyssier1997}.
Then, the pressure built in the high-density core of the halo balances the gravitational
pull and one obtains a static profile in physical coordinates $\vr = a \vx \propto \tau^2 \vx$.
This implies the following relation between the exponents $\alpha$ and $\gamma$
\beq
\alpha = \frac{6}{\gamma} - 2  , \;\;\; 1 < \gamma < 3 , \;\;\; \mbox{hence} \;\;\;
0 < \alpha < 4 .
\label{eq:alpha-gamma}
\eeq
Then, the density contrast reads in physical coordinates
\beq
\delta(r,\tau) = \tau^6 \left( \frac{r}{x_\star} \right)^{-\gamma}
= a^3 \left( \frac{r}{x_\star} \right)^{-\gamma} ,
\label{eq:delta-r}
\eeq
and $\bar\rho \delta(r,\tau)$ is independent of time.
The lower bound $\gamma>1$ corresponds to the fact that for shallower slopes
the core does not converge to a static profile. The mass that keeps collapsing at large
radii at later times is too large and cannot be stabilized, so it continuously
redistributes matter down to the center and the density at a given physical radius
keeps growing with time.
The upper bound $\gamma < 3$ corresponds to the limit of a finite collapsed mass
with negligible or no matter at outer radii; then,
$\alpha=0$ and no more comoving shells turn around, i.e. decouple from the background
cosmological flow and start collapsing, falling towards the central overdensity.
As  we are not interested in the formation of the matter
overdensity itself, we could extend the range of $\gamma$ to $0<\gamma<3$.

The profile (\ref{eq:delta-self-similar}) is sufficient for our purposes, since we are not
interested in building an exact solution to the gravitational collapse of matter
overdensities
\footnote{One can derive exact self-similar solutions of the Newtonian gravitational collapse,
for both collisional and collisionless matter
\cite{Fillmore:1984aa,Bertschinger:1985ab,Teyssier1997}.
The self-similarity means that the nonlinear density, velocity, and pressure profiles
at different times are identical up to a rescaling of the radial coordinates and of the
characteristic density, velocity and pressure.
This symmetry allows one to transform the 2D problem, which involves partial differential
equations over time and radius, into a 1D problem, which involves ordinary differential
equations over a radial coordinate. This enables detailed analytical studies.
The profile (\ref{eq:delta-self-similar}) is a simple approximation to such solutions,
where we extend to all radii the power-law behavior of the density contrast that is
only reached in the nonlinear core of the exact solutions. In all cases
for $x \gg x_s$ we simply recover the background density $\bar\rho$ for the density
$\rho$.}.
Instead, we only wish to study how the scalar field reacts to the formation
of matter overdensities. The power-law form (\ref{eq:delta-self-similar}) allows us
to derive explicit analytical results for a realistic range of density profiles,
parametrized by the exponent $\gamma$.
Of course, if we compare the K-mouflage cosmology with a reference
Einstein-de Sitter cosmology, the matter density profile would be slightly modified
by the fifth force mediated by the scalar field and it would no longer remain self-similar,
even if the initial conditions were power laws (a cosmological constant also breaks
the self-similarity as the scale factor is no longer a power law of time).
However, as explained above, in this paper
we neglect the backreaction of the scalar field and the fifth force. This is consistent
with realistic scenarios, as observations such as Big Bang Nucleosynthesis,
Cosmic Microwave Background and galaxy surveys constrain $\beta \lesssim 0.1$
and the fifth force not to surpass Newtonian gravity.
In any case, the power-law profile (\ref{eq:delta-r}) is only used for computational
convenience, to illustrate general behaviors.
Then, in the K-mouflage case it is understood as a model for the
full density profile including the effect of the fifth force (which
means that the initial condition would be slightly different). Indeed,
in this paper we do not solve the dynamics of the matter, which is
treated as external given data.

Thanks to the simple form of the profile (\ref{eq:delta-self-similar}), we can perform
the integrations in Eq.(\ref{eq:phi-linear}) and we obtain
\begin{widetext}
\beqa
&& \mbox{for} \;\;\;    0<x<\tau :   \nonumber  \\
&& \varphi_L(x,\tau) = \frac{12 \beta x_\star^{\gamma}}{(\gamma-2)(\gamma-3)}
\tau^{\alpha\gamma-2} x^{2-\gamma} + \frac{6\beta x_\star^{\gamma}}{\Gamma(\gamma-1)}
\left\{ - (\alpha\gamma+1) \Gamma(\gamma-5) \tau^{\alpha\gamma-4} x^{4-\gamma}
\left[ (\gamma-5) \frac{\tau}{x} \left( {_2}F_1(1,1-\alpha\gamma;5-\gamma;-\frac{x}{\tau})
\right. \right. \right. \nonumber \\
&& \left. \left. \left.
- {_2}F_1(1,1-\alpha\gamma;5-\gamma;\frac{x}{\tau}) \right)
+ {_2}F_1(1,1-\alpha\gamma;6-\gamma;-\frac{x}{\tau})
+ {_2}F_1(1,1-\alpha\gamma;6-\gamma;\frac{x}{\tau}) \right]
+ \frac{\pi \Gamma(\alpha\gamma) (\alpha\gamma+1)}
{\Gamma(\alpha\gamma+4-\gamma) \sin(\gamma\pi)}
\tau^{\alpha\gamma-\gamma+1} \right. \nonumber \\
&& \left. \times x^{-1} \left[ (1+\frac{x}{\tau})^{\alpha\gamma+3-\gamma}
- (1-\frac{x}{\tau})^{\alpha\gamma+3-\gamma} - (\alpha\gamma+5-\gamma)
\left(  (1+\frac{x}{\tau})^{\alpha\gamma+4-\gamma}
- (1-\frac{x}{\tau})^{\alpha\gamma+4-\gamma} \right) \right]
\right\} ,
\label{eq:phi-self} \\
&& \mbox{and for} \;\;\;   x>\tau :  \nonumber \\
&& \varphi_L(x,\tau) = \frac{6 \beta x_\star^{\gamma}}{\alpha\gamma(\gamma-2)(\gamma-3)}
\tau^{\alpha\gamma-2} x^{2-\gamma} \left\{ (3-\gamma) \frac{\tau}{x}
\left[ {_2}F_1(1,\gamma-2;\alpha\gamma+2;\frac{\tau}{x})
- {_2}F_1(1,\gamma-2;\alpha\gamma+2;-\frac{\tau}{x}) \right] \right. \nonumber \\
&& \left. + {_2}F_1(1,\gamma-3;\alpha\gamma+2;\frac{\tau}{x})
+ {_2}F_1(1,\gamma-3;\alpha\gamma+2;-\frac{\tau}{x}) - 2 \right\} .
\label{eq:phi-self-outer}
\eeqa
\end{widetext}

The solution is not analytic at $x=\tau$.
This explicitly shows the critical role played by the horizon, $x=\tau$, which is
expected on general grounds. Indeed, we typically expect the scalar field to relax
inside the horizon, where information has time to propagate, but not beyond the horizon.

\begin{figure}
\begin{center}
\epsfxsize=8. cm \epsfysize=6 cm {\epsfbox{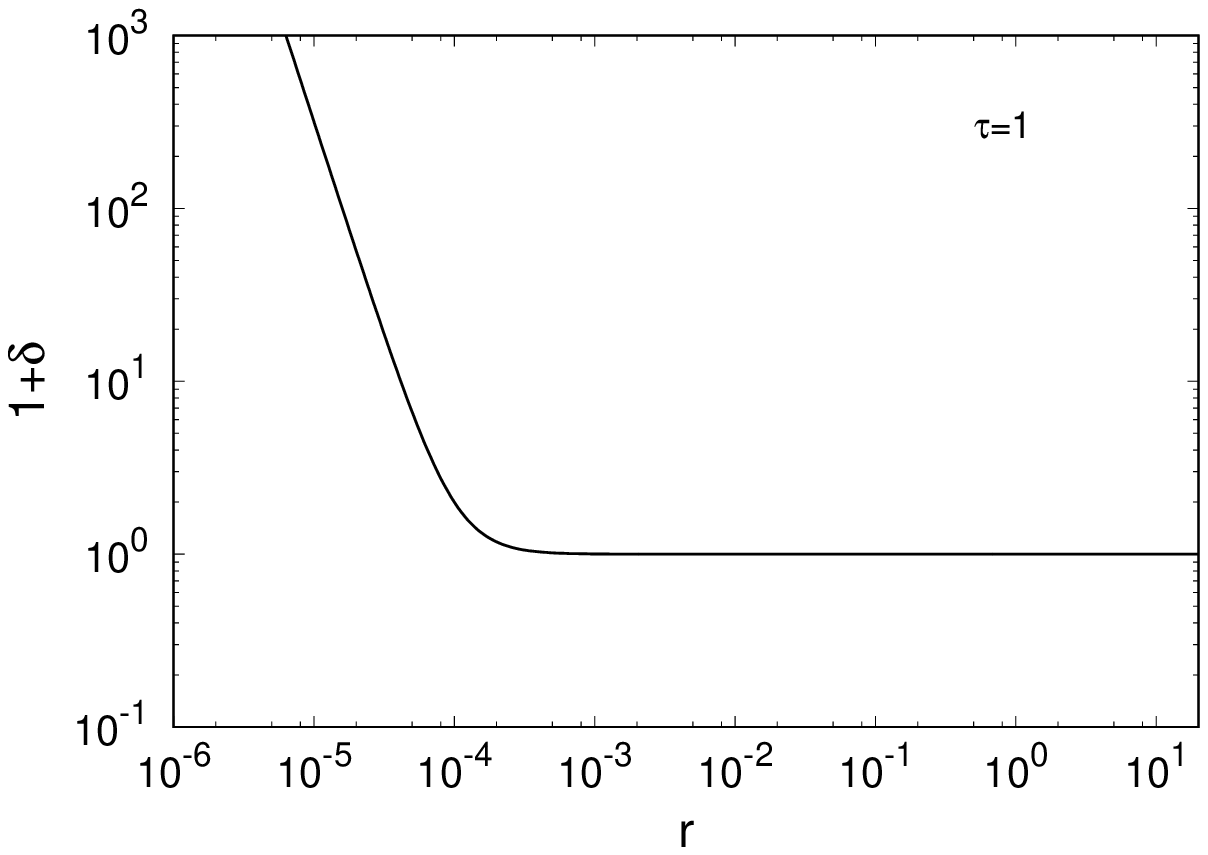}}
\epsfxsize=8. cm \epsfysize=6 cm {\epsfbox{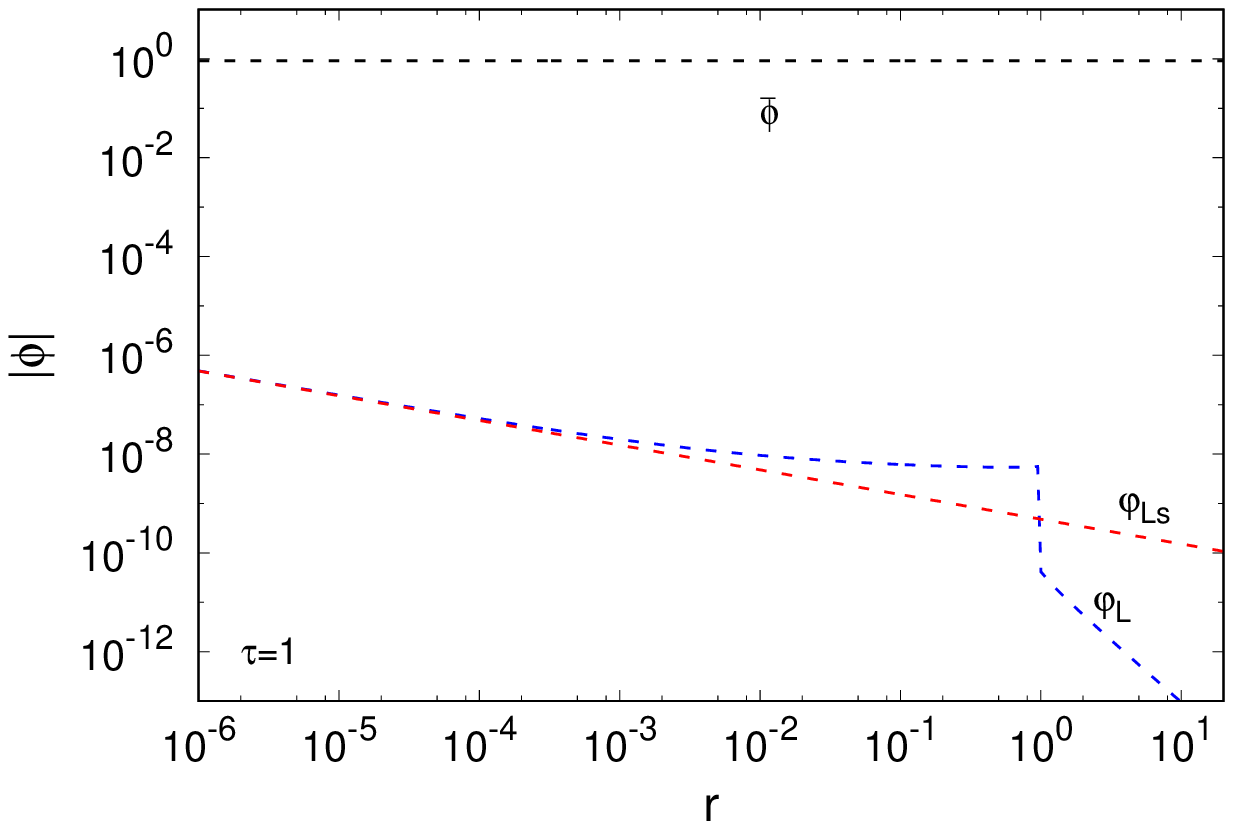}}
\end{center}
\caption{
{\it Upper panel:} matter density contrast $\delta$ today, at $\tau=1$.
{\it Lower panel:} background scalar field $\bar\phi$, linear deviation $\varphi_L$
obtained for constant $K'=1$ from Eqs.(\ref{eq:phi-self}) and (\ref{eq:phi-self-outer}),
and quasistatic solution $\varphi_{Ls}$ from Eq.(\ref{eq:phis-self}).
In all figures in this paper, solid lines correspond to positive values and dashed lines
to negative values, and on logarithmic scales we plot the absolute values.
}
\label{fig_delta}
\end{figure}

On large superhorizon radii, we obtain
\beq
x \gg \tau : \;\;\; \varphi_L \sim \beta \tau^{6-2\gamma} (x/x_\star)^{-\gamma} .
\label{eq:phi-outer}
\eeq
This goes to zero at large radius for all $\gamma > 0$, which shows that
we can indeed expand the range (\ref{eq:alpha-gamma}) to $0<\gamma<3$
for the validity of the linear solution.
On small subhorizon radii, expanding Eq.(\ref{eq:phi-self}) in  $x/\tau$,
we obtain
\beqa
x \ll \tau : && \varphi_L(x,\tau) = \varphi_{Ls}(x,\tau)
\left[ 1 + \left(\frac{x}{\tau}\right)^2 + ... \right] \nonumber \\
&& \;\;\; + \varphi_{Ls}(x=\tau,\tau)  \left[ 1 + \left(\frac{x}{\tau}\right)^2
+ ... \right] ,
\label{eq:phi-self-expand}
\eeqa
where the dots stand for higher orders in $(x/\tau)^2$, and
we omitted  numerical factors except for the first term.
We introduced the leading term $\varphi_{Ls}$, given by the first term in
Eq.(\ref{eq:phi-self}),
\beqa
\varphi_{Ls}(x,\tau) & = & \frac{12 \beta}{(\gamma-2)(\gamma-3)} \frac{x^2}{\tau^2}
\delta(x,\tau) \nonumber \\
& = & \frac{12 \beta}{(\gamma-2)(\gamma-3)} r^2
\left( \frac{r}{x_\star} \right)^{-\gamma} ,
\label{eq:phis-self}
\eeqa
and
\beq
\varphi_{Ls}(x=\tau,\tau) \sim \beta \delta(\tau,\tau)
= \beta \left( \frac{x_s(\tau)}{\tau} \right)^{\gamma} \ll 1 .
\eeq
The term $\varphi_{Ls}(x,\tau) $ actually corresponds to the quasistatic approximation,
where we only keep the spatial derivatives in the Klein-Gordon equation
(\ref{eq:KG-linear-pert}).
Indeed, we can check that it obeys
\beq
\nabla^2 \varphi_{Ls} = 12\beta \frac{\delta}{\tau^2}   \;\;\; \mbox{hence} \;\;\;
\nabla_{\bf r}^2 \varphi_{Ls} = 12\beta \frac{\rho-\bar\rho}{\bar\rho_0} .
\label{eq:quasistatic}
\eeq
Once expressed in terms of the physical radius $r$, it does not depend on time,
as we consider matter overdensities that virialize to static profiles.

We show the density and scalar field profiles at $\tau=1$ in Fig.~\ref{fig_delta}.
Throughout this paper, for the numerical computations we choose the numerical values
\beq
\beta= 0.1 ,  \;\;\;  \gamma = 2.5  ,  \;\;\; x_\star= 10^{-4} .
\label{eq:beta-num}
\eeq
This value of $x_\star$ gives a radius of about $0.6 h^{-1} {\rm Mpc}$ for the
matter overdensity today, which roughly corresponds to the size of galaxy clusters,
but with a steeper slope $\gamma$ to emphasize the nonlinear regime.
We clearly see the discontinuity of $\varphi_L$ at the horizon, $x=\tau=1$,
and the change of slope, from $x^{-\gamma}$ beyond the horizon to
$x^{2-\gamma}$ inside the horizon (we chose a value of $\gamma$ such
that $2-\gamma < 0$). Then, $\varphi_L$ quickly converges to $\varphi_{Ls}$
below the horizon.
On the other hand, the deviation $\varphi_L$ remains much smaller than the background
$\bar\phi$ down to very small radii, so that $\phi_L \simeq \bar\phi$
on most relevant scales.
The discontinuity at the horizon shows that even in the linear case (i.e., when $K'$ is
a constant), the Klein-Gordon equation being a hyperbolic advection equation it can
display shocks. Here, the shock follows the horizon and travels at the
constant speed $dx/d\tau=1$.
This suggests that in such models there could exist a network of discontinuities,
at Hubble distances from matter density caustics, that would produce small kicks
to the velocities of particles that cross these singularities.
A crude estimate for the velocity discontinuity experienced by these particles is
$\Delta v \sim \frac{c^2\beta}{v_0} \Delta\varphi_L$, where $v_0$ is the relative
velocity of the particles. Using $v_0 \simeq c$ and
$\Delta\varphi_L \sim \varphi_{Ls}(x=\tau) \sim \beta \delta(x=\tau)$ gives
$\Delta v \sim c \beta^2 \delta(x=\tau) \ll c$.
In practice, the density profiles do not extend to the horizon; hence we can expect such
velocity kicks to be negligible. However, we leave this issue for other works.

\begin{figure}
\begin{center}
\epsfxsize=8. cm \epsfysize=6 cm {\epsfbox{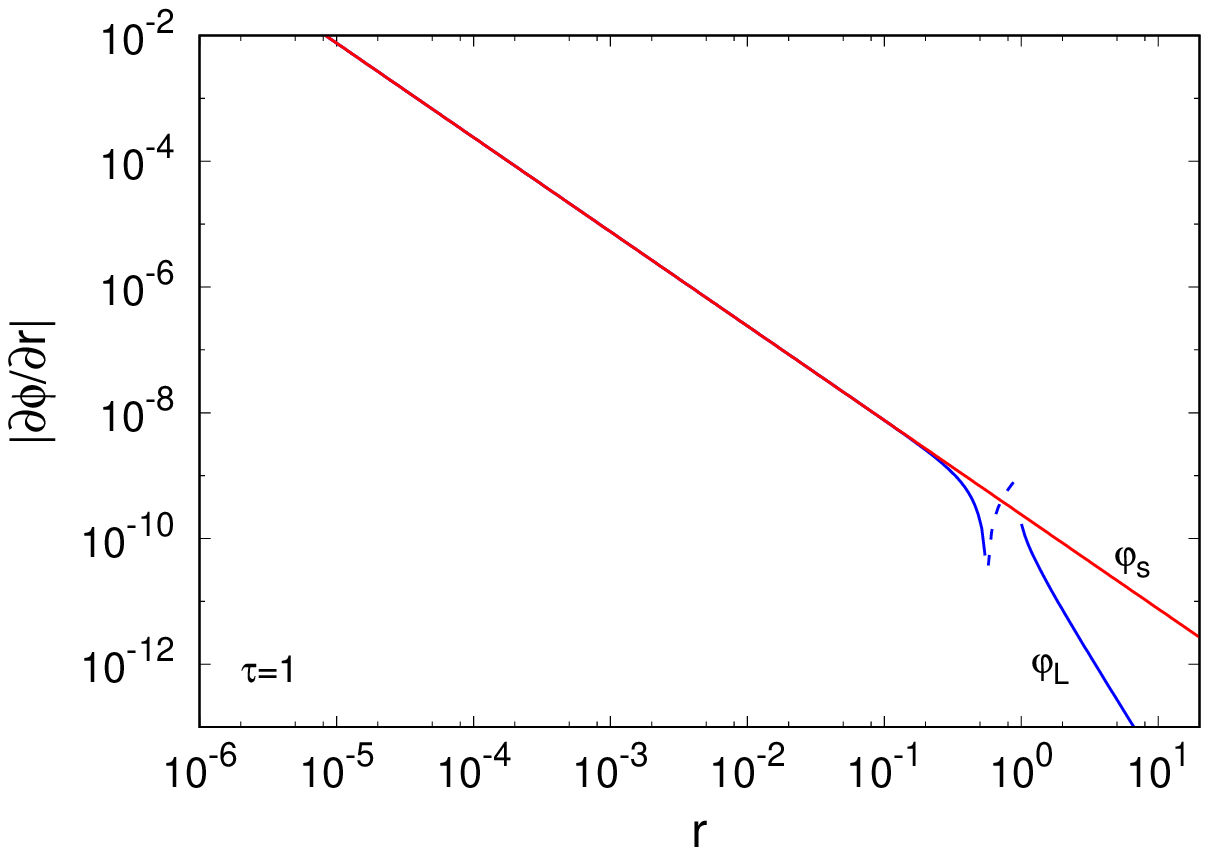}}
\epsfxsize=8. cm \epsfysize=6 cm {\epsfbox{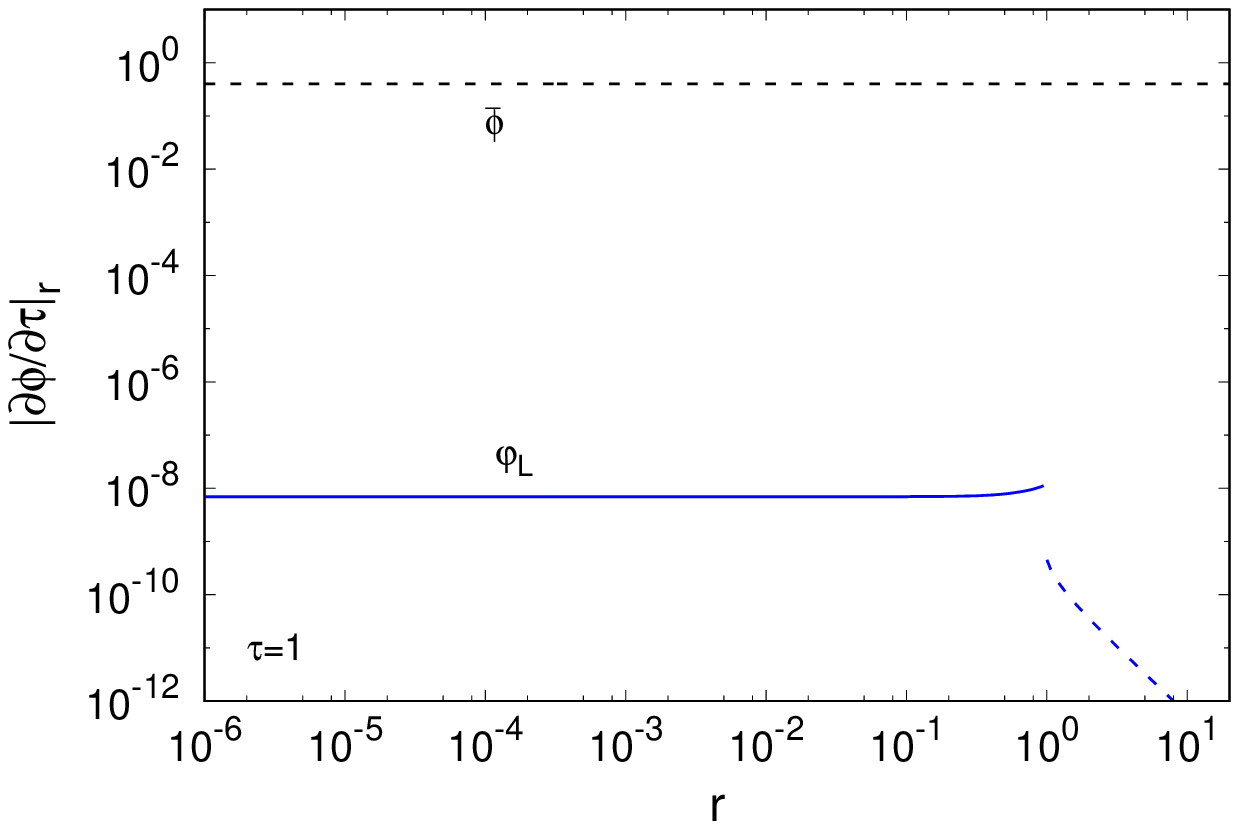}}
\end{center}
\caption{
{\it Upper panel:} radial gradient $\partial_r\phi$, at $\tau=1$, for the linear
and quasistatic solutions.
{\it Lower panel:} time derivative $\partial_\tau\phi |_{r}$,
at fixed physical radius $r$, for the background $\bar\phi$
and the linear deviation $\varphi_L$.
}
\label{fig_dphi_L}
\end{figure}

We can check that the spatial gradients of the exact solution (\ref{eq:phi-self-expand})
are governed by the quasistatic solution at small radii because $\gamma > 0$,
\beq
x \ll \tau : \;\;\; \nabla \varphi_L \simeq \nabla\varphi_{Ls} + \varphi_{Ls}(\tau,\tau)
\frac{x}{\tau^2}
\simeq \nabla\varphi_{Ls} \propto x^{1-\gamma} .
\label{eq:gradient}
\eeq
However, the Poisson equation (\ref{eq:quasistatic}) only defines $\varphi_{Ls}$
up to a constant, if we do not add boundary conditions at large radii.
The explicit solution (\ref{eq:phi-self}) shows that such a term is indeed generated
and can be explicitly calculated.
It becomes time dependent, following the slowly evolving matter overdensity,
and its magnitude is of order $\varphi_{Ls}(x=\tau,\tau)$, as may be expected
(since there are no other scales in the problem).

The partial time derivative at fixed comoving radius $x$ is
\beqa
x \ll \tau : \;\;\; \partial_\tau\varphi_L & \simeq &
\partial_\tau\varphi_{Ls}
+ \frac{d}{d\tau} \varphi_{Ls}(x=\tau,\tau) \nonumber \\
& \sim & \tau^{\alpha\gamma-3} x^{2-\gamma} + \tau^{\alpha\gamma-1-\gamma} .
\eeqa
Throughout this paper, $\partial_\tau\phi = \left. \partial_\tau\phi \right|_x$
stands for the partial time derivative at fixed comoving radius $x$, whereas
we use the subscript ``r'' as in $\left. \partial_\tau\phi \right|_r$
and in Eq.(\ref{eq:time-derivative-r}) below to denote the partial time derivative
at fixed physical radius $r$.
For small radii, $x \ll \tau$, for $\gamma<2$ it is dominated by the second term
and converges to a nonzero value, whereas for $\gamma>2$ it is governed by the
first term and goes to infinity.
Comparing with Eq.(\ref{eq:gradient}), we can see that spatial gradients dominate
over time derivatives at small radii if $\gamma > 1$,
\beq
\gamma > 1 : \;\;\; | \nabla \varphi_L | \gg | \partial_{\tau} \varphi_L |  \;\;\;
\mbox{for} \;\;\; x \ll \tau .
\label{eq:spatial-time-derivatives}
\eeq
For shallower density profiles, $\gamma < 1$, the time derivative associated with
the second term in Eq.(\ref{eq:phi-self-expand}) dominates. This means that
there is no true quasistatic regime in this case, in the sense that the kinetic term
$\chi$ is always dominated by time derivatives.

On small scales, inside the virial radius of the matter overdensity,
it is more appropriate to use the physical coordinates $\{r,\tau\}$
(for simplicity we keep $\tau$ instead of the physical time $t$).
Indeed, we are interested in the impact of the cosmological background
inside nonlinear small-scale structures, such as galaxies or the Solar System,
and we must remove the artificial time dependence due to the use of
comoving coordinates instead of physical coordinates.
In this physical radial coordinate, the density profile after collapse converges
to a constant, as $\bar\rho \delta(r,\tau)$ is independent of time from
(\ref{eq:delta-r}) and $\rho \simeq \bar\rho\delta \gg \bar\rho$.
Then, the time derivative at fixed radius $r$ of the quasistatic solution
is actually zero and the time derivative of the linear solution goes to a constant
at small radii:
\beq
\left. \partial_\tau\varphi_{Ls} \right|_r = 0 ,\;\;\;
\left. \partial_\tau\varphi_L \right|_r \sim \beta x_\star^\gamma \tau^{5-3\gamma}
\;\;\; \mbox{for} \;\;\; x \ll \tau .
\label{eq:time-derivative-r}
\eeq
In particular, this gives at late times inside the matter overdensity
\beq
\tau \sim 1, \;\;\; x\ll \tau : \;\;\; | \partial_\tau\varphi_L |_r
\ll \left| \frac{d\bar\phi}{d\tau} \right| , \;\;\;
\left. \partial_\tau\phi_L \right|_r \simeq \frac{d\bar\phi}{d\tau} ,
\label{eq:time-derivative-r-small}
\eeq
as $x_\star \ll 1$.
We show the spatial and time derivatives in Fig.~\ref{fig_dphi_L}.
We can check that the spatial gradient converges to the quasistatic prediction
on subhorizon scales and that it is much greater than the time derivative
$\partial_\tau \varphi_L |_{r}$.
In agreement with Fig.~\ref{fig_delta} and (\ref{eq:time-derivative-r-small}),
we can check that $| \partial_\tau \varphi_L |_r \ll | d\bar\phi/d\tau|$,
so that the time derivative $\left. \partial_\tau \phi_L\right|_r$ closely follows
the cosmological background on all scales.

For general modified-gravity scenarios involving  an additional scalar field,
the quasistatic approximation is usually understood as
$| \nabla \varphi | \gg | \partial_{\tau} \varphi |$, that is, the spatial gradient of
the scalar field perturbation is greater than its time derivative.
Assuming $\nabla \varphi \sim \varphi/x$ and
$\partial_\tau \varphi \sim {\cal H} \varphi \sim \varphi/\tau$,
one naturally expects this quasistatic regime to hold on subhorizon scales,
$x \ll \tau$. Of course, this also requires that the sound horizon of the scalar field
is of the order of the Hubble radius \cite{Sawicki:2015aa}, i.e. its propagation speed
is of order unity, so that the scalar field has the time to relax on scales $x \ll \tau$.
The validity of this quasistatic regime on subhorizon scales has been
checked for various modified-gravity scenarios, from analytical studies and numerical
simulations, at both the linear \cite{Noller:2014aa}
and nonlinear  \cite{Schmidt:2009af,Bose:2015aa,Winther:2015aa} levels.

The condition $\gamma>1$ in (\ref{eq:spatial-time-derivatives})
shows that for cosmological structures that grow too fast this quasistatic
regime may not be reached, even though the gradients of the scalar field are
already well described by the quasistatic approximation.
In practice, such regimes of fast growth may only occur in transient events, such
as mergings of collapsed halos.
On the other hand, in Cold Dark Matter cosmologies, the variance of the linear matter
density fluctuations behaves as $\sigma^2(x,z) \propto a^2 x^{-(n+3)}$,
where $n$ runs from $1$ to $-3$ from large to small scales, and $n \simeq -2$
on galaxy scales.
This gives for the scale $x_{\rm NL}(\tau)$ that enters the nonlinear regime
\beq
x_{\rm NL}(\tau) \propto \tau^{4/(n+3)} , \;\; \mbox{hence} \;\; \alpha=\frac{4}{n+3} .
\eeq
In the stable-clustering ansatz \cite{Peebles1980},
this gives a slope in the nonlinear regime for the two-point correlation function
\beq
x \ll x_{\rm NL} : \;\;\; \xi(x) \propto x^{-3(n+3)/(n+5)} , \;\;
 \mbox{hence} \;\; \gamma=\frac{3(n+3)}{n+5} .
 \label{eq:stable-clustering}
\eeq
These exponents $\alpha$ and $\gamma$ satisfy the relationship
(\ref{eq:alpha-gamma}).
The stable-clustering ansatz (\ref{eq:stable-clustering}) is not very accurate
\cite{Smith2003},
and in practice it has been replaced by halo models \cite{Cooray2002},
or numerical simulations.
However, it suggests that for $n \leq -2$ and for redshifts $z \gtrsim 2$
the quasistatic condition $| \nabla \varphi | \gg | \partial_{\tau} \varphi |$
may not always be fulfilled as the cosmic web shows a fast buildup.
On the other hand, as the fifth force (i.e. the scalar field gradients) remains
well predicted by the quasistatic approximation and the impact of dark energy
typically becomes negligible at high redshifts, these deviations from the usual
quasistatic condition are unlikely to have important effects.

\begin{figure}
\begin{center}
\epsfxsize=8. cm \epsfysize=6 cm {\epsfbox{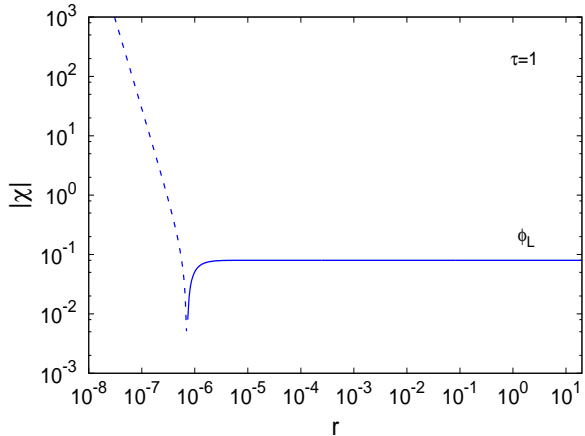}}
\end{center}
\caption{
Kinetic argument $\chi$, at $\tau=1$, for the linear solution $\phi_L$.
}
\label{fig_chi_L}
\end{figure}

The full solution to the Klein-Gordon equation (\ref{eq:KG-linear}) is
$\phi_L=\bar\phi+\varphi_L$.
The background term $\bar\phi$ does not contribute to the spatial gradients
but it contributes to the time derivative.
In particular, for $\gamma<2$ and $\tau \sim 1$ it dominates over the
time derivative $\partial_\tau\varphi_L$ at small radii, and for all $\gamma$
it dominates for $x \sim \tau$.
This means that the spatial gradients $\nabla\phi_L$ dominate over the
time derivative $\partial_\tau\phi_L$ over a smaller range than in
(\ref{eq:spatial-time-derivatives}).
Comparing Eqs.(\ref{eq:phi-bar}) and (\ref{eq:gradient}) we obtain
\beq
\gamma > 1 : \;\;\; |\nabla\phi_L| \gg |\partial_\tau\phi_L| \;\;\; \mbox{for} \;\;\;
x \ll x_{\rm qs}(\tau) ,
\eeq
with
\beq
x_{\rm qs}(\tau) = x_s(\tau) \left( \frac{x_s(\tau)}{\tau} \right)^{1/(\gamma-1)}
\ll x_s(\tau) ,
\label{eq:x-qs-def}
\eeq
where $x_s(\tau)$ is the size of the overdensity, defined in
Eq.(\ref{eq:delta-self-similar}).
As the overdense region must always remain far inside the Hubble radius,
$x_s(\tau) \ll \tau$, we find that $x_{\rm qs} \ll x_s$. Thus, the fully
quasistatic regime, defined as $|\nabla\phi| \gg |\partial_\tau\phi|$,
only applies far inside the overdense region. This is consistent with the fact
that clusters of galaxies are not screened, as found in \cite{Brax:2015aa}.

We show the kinetic argument $\chi$, defined in Eq.(\ref{eq:chi-tilde}),
in Fig.~\ref{fig_chi_L}.
It goes to a constant on large scales, where it is dominated by the background
time derivative, while it grows on small scales, where it follows the spatial gradient
of the quasistatic solution.
We can check that the location of the transition agrees with Eq.(\ref{eq:x-qs-def}),
which gives $x_{\rm qs} \simeq 2 \times 10^{-7}$ at $\tau=1$.
It is far inside the matter overdensity, and the spatial gradient has already
converged to the quasistatic approximation.

In the outer parts, $x_{\rm qs} \ll x \ll x_s$, where the density contrast is
already much greater than unity and the density profile has converged to its static
limit in physical coordinates, the scalar field $\phi_L$ has not yet converged to
a full quasistatic regime in the sense that $|\nabla\phi_L| \ll |\partial_\tau\phi_L|$.
However, its spatial gradients have already converged to the quasistatic prediction,
in fact as soon as $x \ll \tau$, that is, far beyond the size of the overdensity.
For $\gamma < 2$, the value at the center of the scalar field is dominated
by the background,
\beq
\gamma < 2 : \;\;\; \phi_L(0) = \varphi_L(0) + \bar\phi \simeq \bar\phi
= - 4 \beta\ln\tau ,
\label{eq:phi0-drift}
\eeq
whereas for $\gamma > 2$ it is dominated by the quasistatic solution $\varphi_{Ls}$,
which goes to infinity.
In realistic cases, the matter density and the scalar field remain finite inside
collapsed structures and the central value of the scalar field follows the
cosmological drift, in agreement with (\ref{eq:time-derivative-r-small}).

The two conditions $| \nabla\varphi_L | \gg | \partial_\tau\varphi_L |$ and
$| \nabla\phi_L | \gg | \partial_\tau\phi_L |$ define two different quasistatic regimes.
The first condition, which has a greater range of validity, is often used to
study linear perturbations. However, once we take into account nonlinearities
and screening mechanisms, the second condition is more adequate,
as it is a necessary condition for local screening of the fifth force and
for a local analysis, where the computation of the fifth force does not depend
on the cosmological background and the history of the scalar field evolution.

Thus, even in the simple case where the kinetic term $K'$ is constant and the
Klein-Gordon equation is linear, the quasistatic limit is not so simple.
As expected, spatial gradients converge to the quasistatic prediction as soon as
$x \ll \tau$, i.e. inside the horizon.
Indeed, as the scalar field propagation speed is unity, it has time to relax
and follow the slow cosmological evolution of the density field on subhorizon scales.
The same convergence to the quasistatic limit was found in studies of
modified-gravity models that display the Vainshtein mechanism,
which also involves a wave equation and a similar derivative screening
\cite{Winther:2015aa}.

However, time derivatives remain dominant
down to the much smaller radius $x_{\rm qs}$, far inside the nonlinear overdense region,
where they are dominated by the cosmological background.
If $\gamma <2$ and the quasistatic solution is finite at the center,
which is the case in realistic matter overdensities, the value of the scalar field
at the center remains governed by the cosmological background.
This shows that the quasistatic approximation predicts the spatial gradients,
hence the fifth force, with  great accuracy on all subhorizon scales.
However, the scalar field does not decouple from the cosmological background,
except at the very center for steep density profiles in the particular case where
it becomes infinite.
This also shows that both the nonlinear transition and the quasistatic regime
of the scalar field differ from their counterparts for the matter density field.

\subsection{Static compact object}

The power-law profiles (\ref{eq:delta-self-similar}) allowed us to study the
evolution of the scalar field for a variety of matter density profile exponents
and for cosmological structures that keep growing with time.
It is also interesting to consider small-scale structures that no longer grow,
with a constant matter density. This corresponds to compact objects
such as stars, the Solar System, or an isolated galaxy.
Thus, we consider the top-hat density profiles
\beq
\tau > \tau_* : \;\;\; \delta(\vx,\tau)
= \theta(\frac{r_*}{\tau^2}-x) \frac{\tau^6}{\tau_*^6}  \;\; \mbox{with} \;\;
r_* \ll \tau_*^3 ,
\label{eq:delta-top-hat}
\eeq
and $\delta=0$ for $\tau < \tau_*$.
This corresponds to matter overdensities that form at time $\tau_*$, with a fixed
physical radius $r_*$ and density $\rho_* \sim \bar\rho(\tau_*)$, so that
$\delta$ grows as $a^3$ at later times.
The condition $r_* \ll \tau_*^3$ means that the structure is far inside the Hubble radius
at formation time.
From Eq.(\ref{eq:phi-linear}) we obtain the solution as
\beqa
&& \varphi_L(x,\tau) = \frac{24\beta}{\pi\tau\tau_*^6} \int_{\tau_*}^\tau d\tau'
\tau'^7 \int_0^{\infty} dk \frac{\sin(kx)}{kx} \nonumber \\
&& \times [ \sin(k \frac{r_*}{\tau'^2})
- k \frac{r_*}{\tau'^2} \cos(k \frac{r_*}{\tau'^2}) ] \nonumber \\
&& \times [ n_1(k\tau') j_1(k\tau) - j_1(k\tau') n_1(k\tau) ] .
\eeqa
We could not derive a simple explicit expression for the profile of the scalar field,
but we can obtain the value at the center, which at leading order over $r_*$ reads as
\beq
\varphi_L(0,\tau) \simeq -6\beta \frac{r_*^2}{\tau_*^6} .
\label{eq:phi0-top-hat}
\eeq
Thus, as for the self-similar profiles in Eq.(\ref{eq:phi0-drift}),
we find that the scalar field closely follows the cosmological drift with
$\phi_L(0) \simeq \bar\phi$.

We can now check that $|\varphi_L(0)| \sim |\varphi_{Ls}(x=\tau,\tau)|$, in agreement
with the expansion (\ref{eq:phi-self-expand}) that we explicitly derived
for the power-law profiles.
For the top-hat profile (\ref{eq:delta-top-hat}), the quasistatic solution
that corresponds to Eq.(\ref{eq:phis-self}), normalized to zero at the center,
reads as
\beqa
0 < x < \frac{r_*}{\tau^2} : && \varphi_{Ls}(x) = \frac{2\beta \tau^4 x^2}{\tau_*^6} ,
\nonumber \\
x > \frac{r_*}{\tau^2} : && \varphi_{Ls}(x) = \frac{6\beta r_*^2}{\tau_*^6}
- \frac{4\beta r_*^3}{\tau^2 \tau_*^6 x} .
\eeqa
This gives $\varphi_{Ls}(x=\tau,\tau) \simeq 6 \beta r_*^2/\tau_*^6$,
which is of the same order of magnitude as (\ref{eq:phi0-top-hat}).
This confirms the general behaviors found in section~\ref{sec:self-similar}
for the power-law profiles.

\section{Nonlinear kinetic term}
\label{sec:nonlinear}

\subsection{Screening radius and quasistatic solution}
\label{sec:2-step}

We will now consider the impact of the nonlinear K-mouflage screening mechanism.
As recalled in the introduction, the effects of the nonlinearity of the kinetic function
$K$ on the cosmological background and on small-scale astrophysical structures
are independent as they are related to the two separate regimes $\chi\to+\infty$
and $\chi\to-\infty$.
The nonlinear impact on the cosmological background is simple to analyze
\cite{Brax:2014a,Brax:2015aa},
and follows from the nonlinear ordinary differential equation
(\ref{eq:KG-background}).
In this paper, we are interested in the nonlinearities that occur in small-scale
high-density environments, associated with large negative $\chi$, that also screen
the fifth force in the Solar System. Therefore, we keep $K'=1$ for positive $\chi$
and focus on the nonlinear screening associated with large spatial gradients of
the scalar field. More precisely, we consider the case where the kinetic function $K'$
remains constant and equal to unity over all $\chi>\chi_{\rm sc}$,
with $-\chi_{\rm sc} \gg 1$,
\beq
\chi < \chi_{\rm sc} : \;\;\; K'(\chi) \gg 1  , \;\;\;
\chi > \chi_{\rm sc} : \;\;\; K'(\chi) = 1  .
\label{eq:chi-sc-def}
\eeq
The threshold $\chi_{\rm sc}$ determines the boundary $x_{\rm sc}(\tau)$
of the screened region, where $K' \gg 1$ and the fifth force is damped by
the K-mouflage screening mechanism,
\beq
\chi = \chi_{\rm sc} \;\;\; \mbox{at} \;\;\; x = x_{\rm sc}(\tau) .
\label{eq:s-sc-def}
\eeq

It is useful to first consider a generalized quasistatic solution.
Indeed, as for the linear case studied in section~\ref{sec:linear}, we can anticipate
that at small radii the radial profile of the scalar field will be determined by the
quasistatic solution. On the other hand, we also expect the time derivative to remain
set by the cosmological background, at least on large scales.
Thus, we define the generalized quasistatic solution $\varphi_s$ by
\beq
\nabla ( K' \nabla \varphi_s ) = 12 \beta \frac{\delta}{\tau^2} , \;\;\;
\chi = \frac{1}{2\tau^4} \left[ \left( \frac{d\bar\phi}{d\tau} \right)^2
- (\nabla\varphi_s)^2 \right] .
\label{eq:qsK}
\eeq
This generalizes to the nonlinear case the previous equation (\ref{eq:quasistatic}).
As in the linear case, we separate the source $\delta$ associated with the
matter overdensity from the unit factor of the term $(1+\delta)$,
which is related to the mean cosmological background, and we only keep the
spatial derivatives in the Klein-Gordon equation (\ref{eq:KG-3}), which becomes
a nonlinear Poisson equation. However, we keep the time derivative in the kinetic
argument $\chi$, using its background value.
This ensures that we recover the right limit for $K'$ on large scales.

For a spherically symmetric overdensity, integrating this nonlinear Poisson equation
once, we obtain
\beq
K' \frac{d\varphi_s}{dx} = \frac{12\beta}{x^2\tau^2} \int_0^x dx \, x^2\delta
\label{eq:phis-screening}
\eeq
with
\beq
\chi_s \simeq - \frac{1}{2\tau^4} \left( \frac{d\varphi_s}{dx} \right)^2 \;\;\;
\mbox{for} \;\;\; x \ll x_{\rm qs} ,
\label{eq:chis}
\eeq
on small scales in the spatial domain, where $\chi$ is dominated by the spatial gradient.
At large radii $x>x_{\rm sc}$, where $\chi>\chi_{\rm sc}$, we have $K'=1$
and we obtain the explicit expression
\beq
x >x_{\rm sc} : \;\;\;
\frac{d\varphi_s}{dx} = \frac{12\beta}{x^2\tau^2} \int_0^x dx \, x^2\delta ,
\eeq
independently of the nonlinear behavior at smaller radii.
For the self-similar density profile (\ref{eq:delta-self-similar}) this gives
\beq
\frac{d\varphi_s}{dx} = \frac{12\beta}{3-\gamma} x_\star^{\gamma}
\tau^{\alpha\gamma-2} x^{1-\gamma} \sim \beta \frac{x}{\tau^2} \delta
\eeq
and
\beq
\chi_s = - \frac{1}{2} \left( \frac{12\beta}{3-\gamma}\right)^2 x_\star^{2\gamma}
\tau^{2\alpha\gamma-8} x^{2-2\gamma} \sim - \beta^2 \frac{x^2}{\tau^8} \delta^2 ,
\eeq
which coincide with the results obtained from (\ref{eq:phis-self})
in the case of the standard kinetic term.

In this paper, we investigate whether the nonlinearity of the kinetic function
can decouple small-scale structures from the cosmological background.
Therefore, we consider the case $\gamma > 1$, where the gradient $d\varphi_s/dx$
and the magnitude of the argument $\chi_s$ of the kinetic function grow at smaller radii,
so that the core of the overdensity enters the nonlinear screening regime.
The threshold $\chi_{\rm sc}$ is reached by $\chi$ at the radius $x_{\rm sc}(\tau)$,
given by
\beq
x_{\rm sc}(\tau) = x_{\rm qs}(\tau) \left( \frac{12\beta}
{(3-\gamma)\sqrt{-2\chi_{\rm sc}}\tau^3} \right)^{1/(\gamma-1)}  .
\label{eq:x-sc}
\eeq
Since $-\chi_{\rm sc} \gg 1$, at late times $\tau \sim 1$ the screening radius
$x_{\rm sc}$ is far inside the quasistatic region $x_{\rm qs}$.
However, at early times this is not the case anymore as  $x_{\rm sc}/x_{\rm qs}$
grows and  becomes of order unity at the time $\tau_{\rm sc}$ given by
\beq
\tau_{\rm sc} = \left( \frac{12\beta}{(3-\gamma)\sqrt{-2\chi_{\rm sc}}} \right)^{1/3} ,
\label{eq:tau-sc}
\eeq
which is independent of $x_\star$.
This provides a small-time cutoff, as for earlier times the quasistatic approximation
no longer holds up to $x_{\rm sc}$ given by Eq.(\ref{eq:x-sc}).
Using the relationship (\ref{eq:alpha-gamma}), we can see from the expression
(\ref{eq:x-sc}) that $x_{\rm sc} \propto \tau^{-2}$, that is,
\beq
\tau > \tau_{\rm sc} : \;\;\; x_{\rm sc}(\tau) = \frac{r_{\rm sc}}{\tau^2}
= \frac{r_{\rm sc}}{a} ,
\label{eq:r-sc}
\eeq
where $r_{\rm sc}$ is constant,
\beq
r_{\rm sc} = x_\star  \left( \frac{12\beta x_\star}{(3-\gamma)\sqrt{-2\chi_{\rm sc}}}
\right)^{1/(\gamma-1)}  .
\label{eq:r-sc-def}
\eeq
This means that in physical coordinates the screening radius $r_{\rm sc}$
does not depend on time. This is a direct consequence of the fact that the density
profile (\ref{eq:delta-self-similar}) converges to a static profile in physical
coordinates, in the nonlinear region $\delta \gg 1$.
There, $d\varphi_s/dr$ and $\chi_s$ also converge to a static profile
in physical coordinates, so that the threshold $\chi_{\rm sc}$ corresponds to
a constant physical radius $r_{\rm sc}$.

\subsection{Numerical analysis}
\label{sec:numerical}

\begin{figure}
\begin{center}
\epsfxsize=8. cm \epsfysize=6 cm {\epsfbox{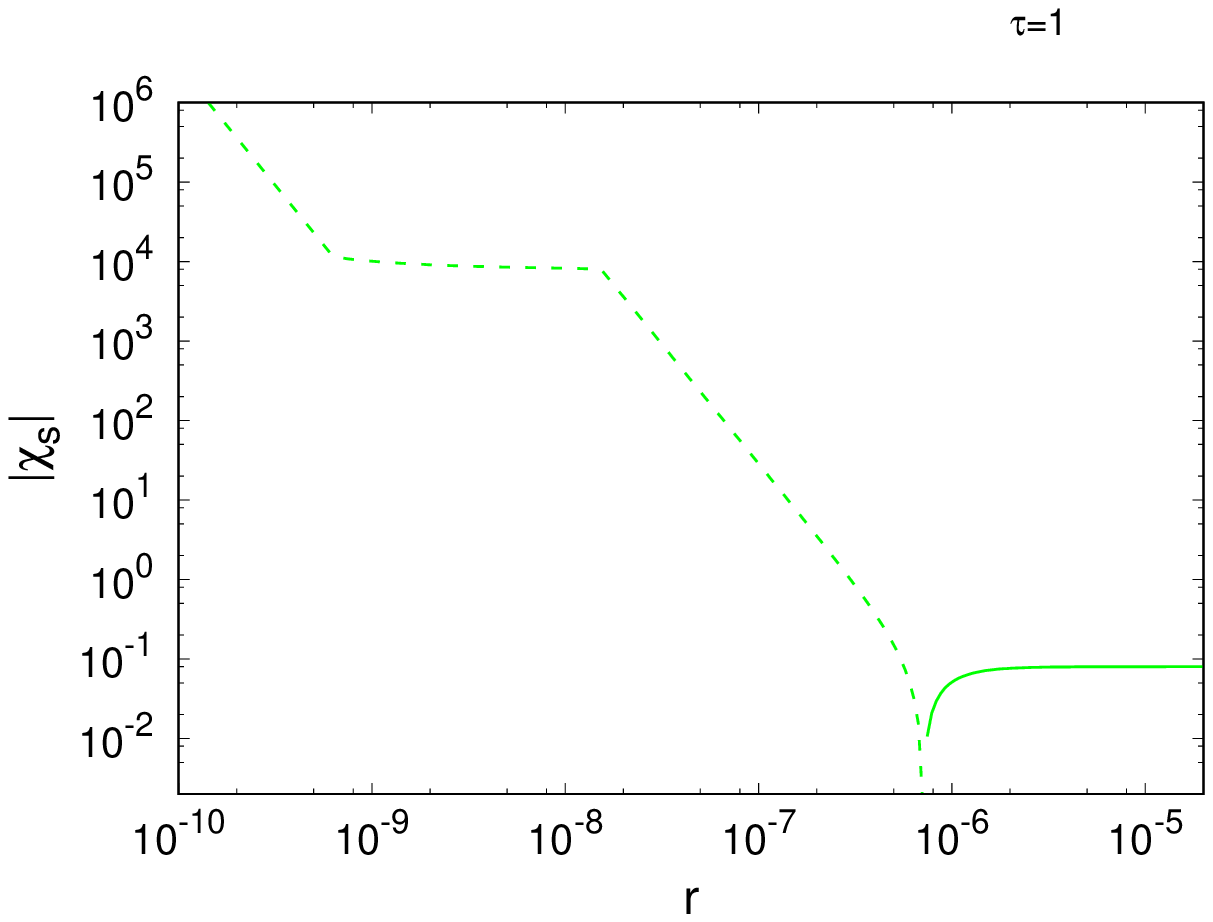}}
\epsfxsize=8. cm \epsfysize=6 cm {\epsfbox{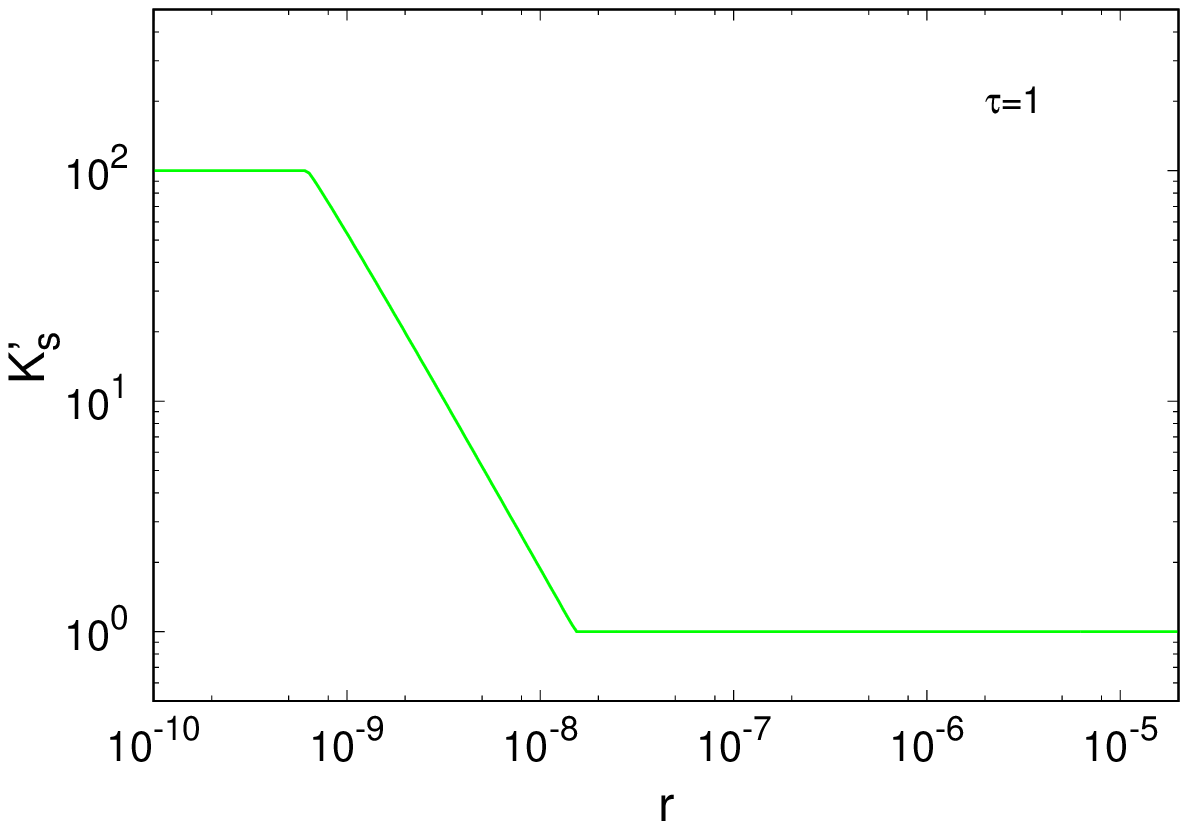}}
\end{center}
\caption{
{\it Upper panel:} kinetic argument $\chi$ for the generalized quasistatic solution
(\ref{eq:qsK}), at $\tau=1$.
{\it Lower panel:} kinetic function $K'$ for the generalized quasistatic solution.
}
\label{fig_chi_s}
\end{figure}

\begin{figure*}
\begin{center}
\epsfxsize=8. cm \epsfysize=6 cm {\epsfbox{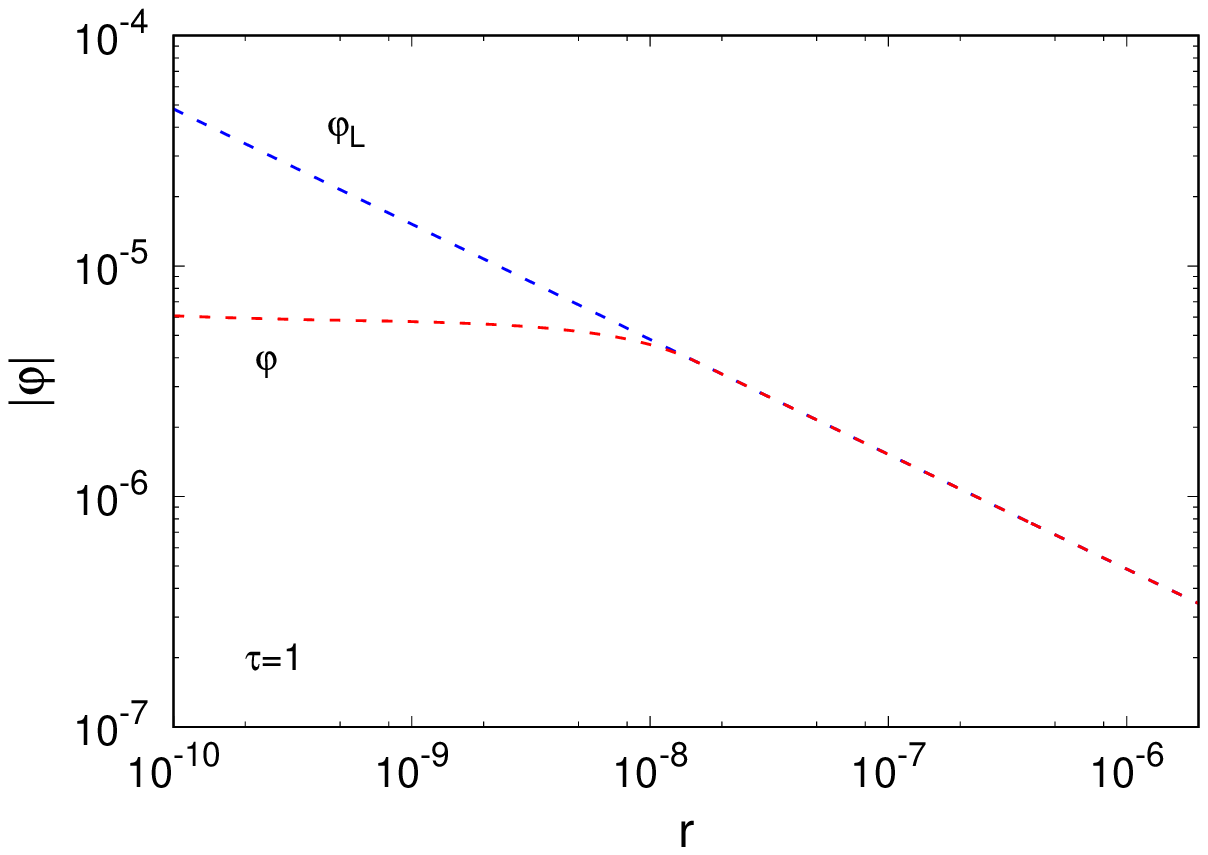}}
\epsfxsize=8. cm \epsfysize=6 cm {\epsfbox{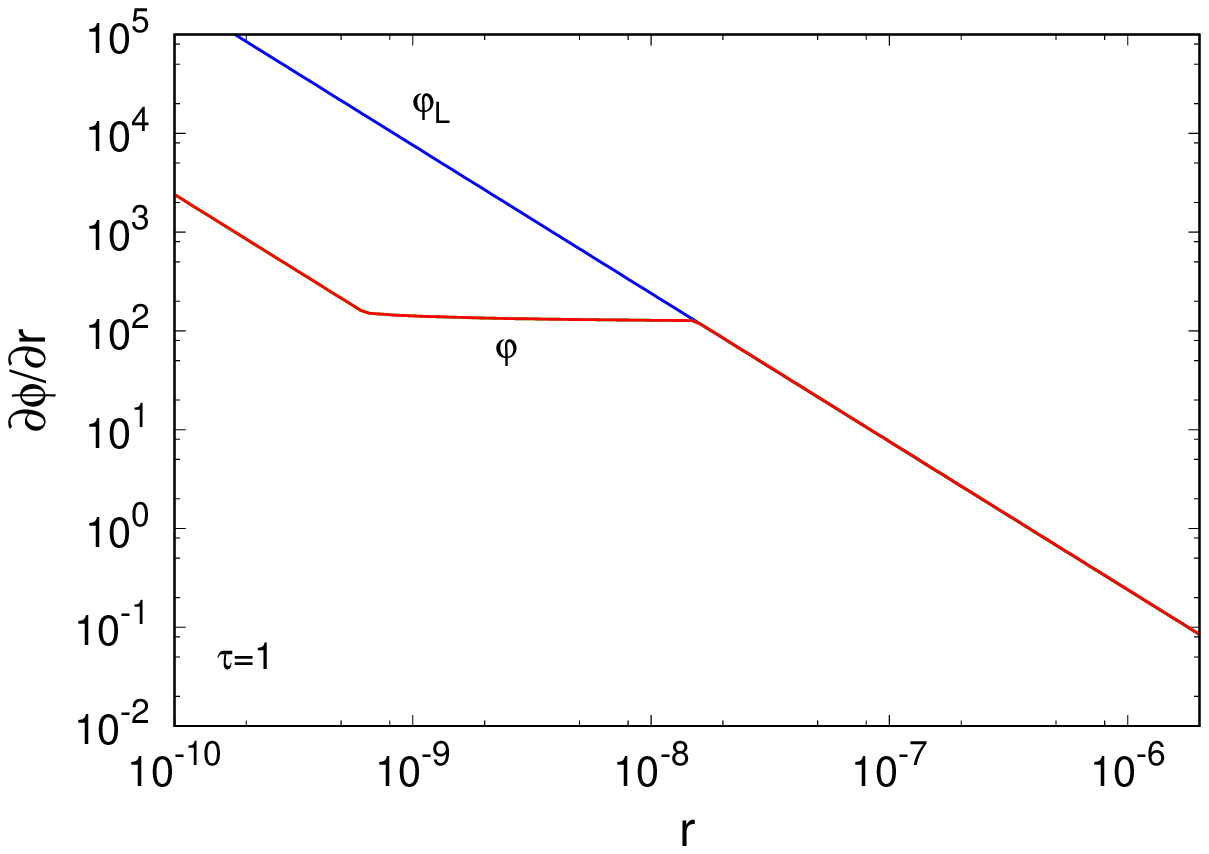}}\\
\epsfxsize=8. cm \epsfysize=6 cm {\epsfbox{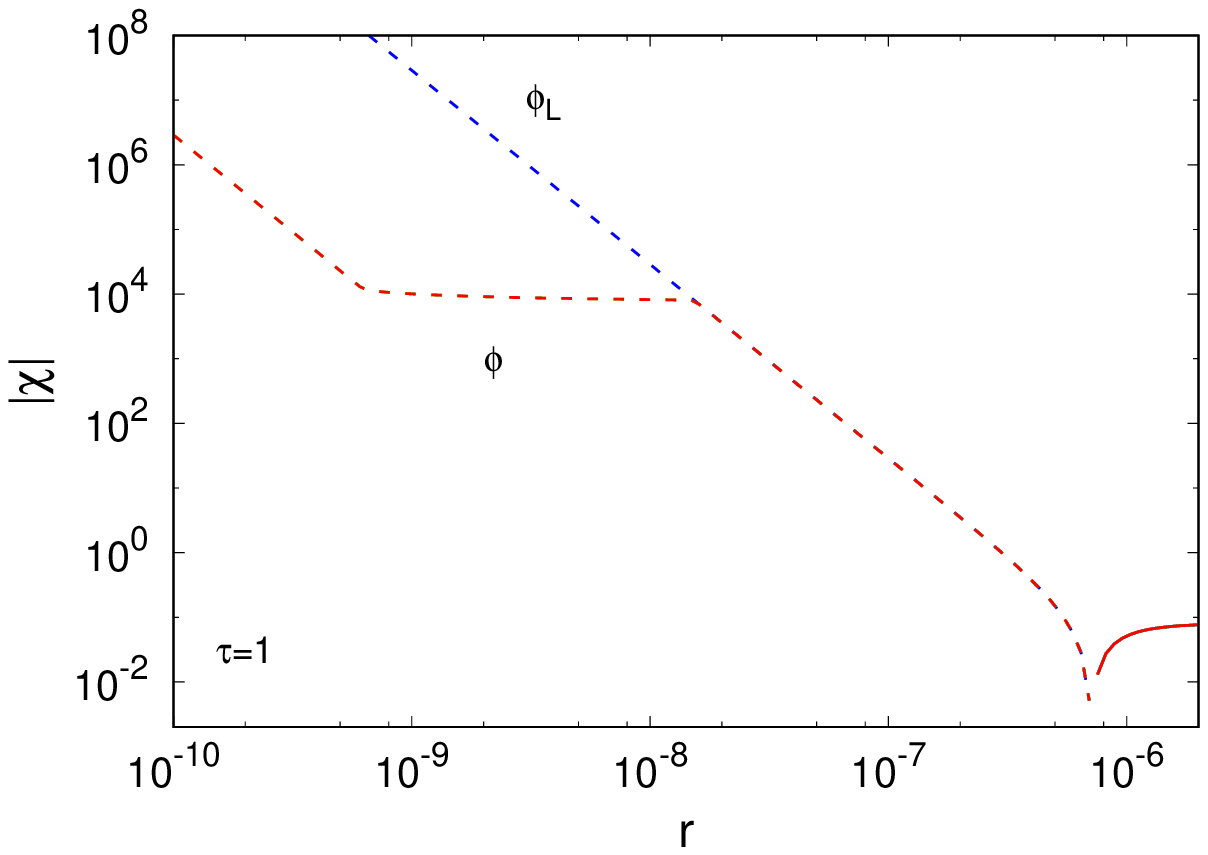}}
\epsfxsize=8. cm \epsfysize=6 cm {\epsfbox{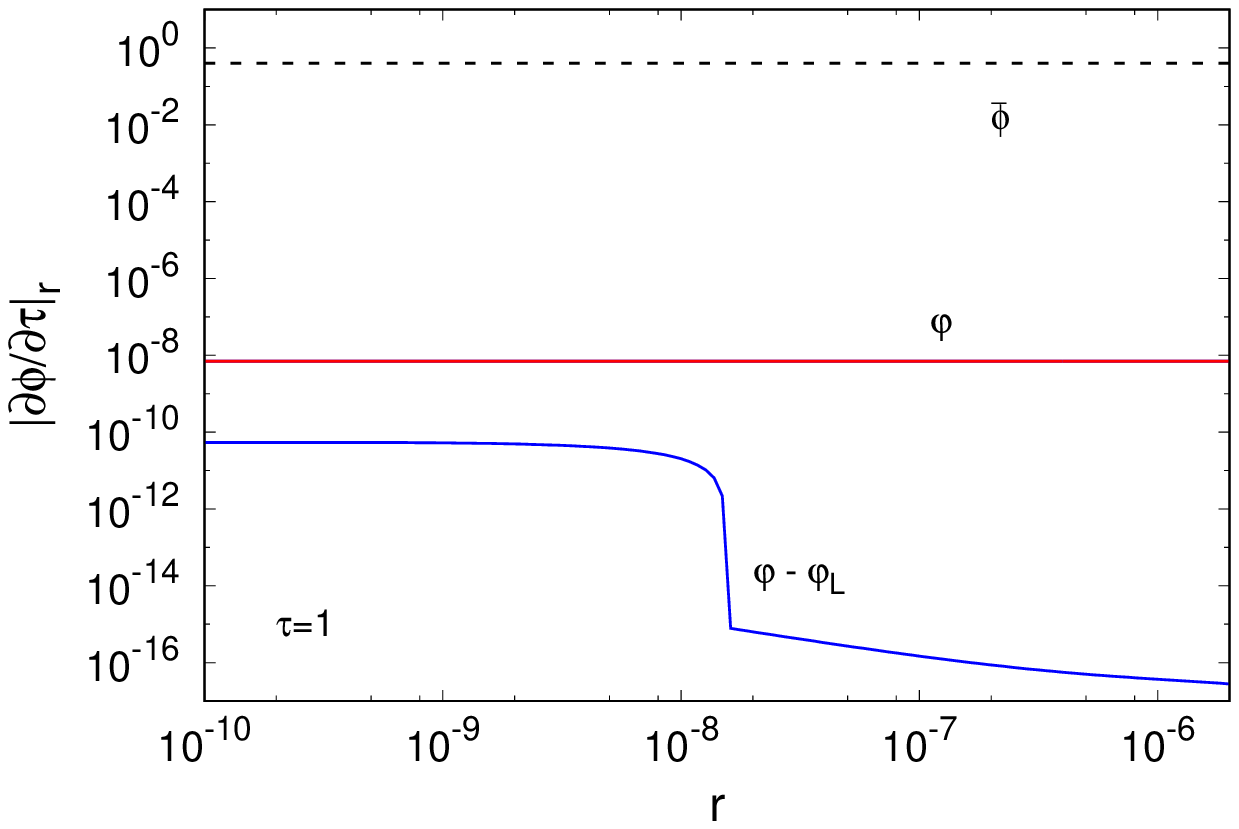}}
\end{center}
\caption{
{\it Upper left panel:} radial profile of the scalar field, at $\tau=1$, for the
nonlinear solution $\phi$ and the linear solution $\phi_L$ associated with $K'=1$.
{\it Upper right panel:} radial gradient $\partial_r\phi$, for the nonlinear
and linear solutions. The nonlinear solution cannot be distinguished from the
generalized quasistatic approximation $\varphi_s$.
{\it Lower left panel:} kinetic argument $\chi$.
{\it Lower right panel:} time derivative $\partial_\tau\phi |_{r}$,
at fixed physical radius $r$, for the background $\bar\phi$ and the
nonlinear deviation $\varphi$, which cannot be distinguished from
the linear deviation $\varphi_L$. We also show the difference
$\varphi - \varphi_L$.
}
\label{fig_dphi}
\end{figure*}

We now perform a numerical analysis of the nonlinear case.
We choose for the kinetic function a simple example of the class (\ref{eq:chi-sc-def}),
with
\beqa
&& \chi < \chi_{\rm sc}-\sigma_{\rm sc} : \;\; K' = K'_{\rm sc} , \nonumber \\
&& \chi > \chi_{\rm sc}+\sigma_{\rm sc} : \;\; K' = 1 ,
\label{eq:Kp-def}
\eeqa
and over the transition range $\chi_{\rm sc}-\sigma_{\rm sc} < \chi < \chi_{\rm sc}+\sigma_{\rm sc}$
we choose for $K'(\chi)$ the cubic polynomial that goes from $K'_{\rm sc}$ down to unity
with vanishing derivative at both ends.
This provides a smooth transition of nonzero width $2 \sigma_{\rm sc}$.
For our numerical computations we choose the values
\beq
\chi_{\rm sc} = -10^4 , \;\;\; \sigma_{\rm sc} = 2000 , \;\;\;
K'_{\rm sc} = 100 ,
\label{eq:chis-num}
\eeq
and we again use the power-law profiles (\ref{eq:delta-self-similar}) for the matter density
contrast, with the same parameters (\ref{eq:beta-num}).
This gives in particular for the screening radius (\ref{eq:r-sc-def})
\beq
r_{\rm sc} \simeq 10^{-8} .
\label{eq:r-sc-num}
\eeq
As $\chi_{\rm sc}+\sigma_{\rm sc}<0$, we have $K'=1$ for all $\chi \geq 0$.
Therefore, the cosmological background is still given by Eq.(\ref{eq:phi-bar}) at all times,
and at large radii we expect to recover the linear solution found in section~\ref{sec:self-similar}.

\subsubsection{Generalized quasistatic approximation}
\label{sec:generalized-quasistatic}

We first show in Fig.~\ref{fig_chi_s} the generalized quasistatic solution (\ref{eq:qsK}).
At large radii we have $\chi \simeq \bar\chi > 0$, as $\chi$ is dominated by the
background time derivative, and $K'=1$.
At small radii we have $-\chi \gg 1$ as it follows the growing spatial derivative of the
quasistatic solution.
The higher value of $K'$ leads to smaller values of the spatial gradient
$\partial_r\phi$ and of $\chi$ as compared with the case $K'=1$
(as seen from the plateau in $\chi$ at the transition). This is the K-mouflage
screening mechanism, which damps the fifth force.

The transition between the time and spatial domains occurs at
$x_{\rm qs} \simeq 2 \times 10^{-7}$,
as for the linear solution, in agreement with Eq.(\ref{eq:x-qs-def}) and
Fig.~\ref{fig_chi_L}, as $K'$ remains unity at this radius.
The transition to the nonlinear regime occurs at the smaller radius
$x_{\rm sc} \simeq 10^{-8}$, in agreement with Eq.(\ref{eq:r-sc-num}),
further within the spatial domain.
From the integrated form (\ref{eq:phis-screening}) of the quasistatic nonlinear
Klein-Gordon equation, we can see that $K' \sqrt{-\chi}$ is a smooth function of radii.
It is a power law for our power-law density profiles.
Then, for functions $K'(\chi)$ that display a sharp transition around $\chi_{\rm sc}$,
from $K'_+$ to $K'_-$ as $\chi$ grows from $\chi_+$ to $\chi_-$, with
$\chi_+ \simeq \chi_- \simeq \chi_{\rm sc}$, the argument $\chi$ remains roughly constant,
close to $\chi_{\rm sc}$, while $K'$ decreases from $K'_+$ to $K'_-$.
Thus, the transition is broad over the radial coordinate, with $K'(x)$ behaving
as a power law whereas $\chi$ is almost constant.
We can clearly see this behavior in Fig.~\ref{fig_chi_s}, although the finite width
$\sigma_{\rm sc}$ smoothens and slightly tilts the plateau for $\chi$.
Thus, the nonlinear Poisson equation (\ref{eq:phis-screening}) is self-regularizing.
Jumps or sharp transitions in the underlying kinetic function $K'(\chi)$ do not give rise
to discontinuities or increasingly steep transitions for the scalar field gradient,
which instead remains roughly constant over the transition.

\subsubsection{Nonlinear solution}
\label{sec:nonlinear-solution}

We can expect the exact solution to follow closely the generalized
quasistatic approximation (\ref{eq:qsK}), with a plateau for $\chi$,
and $\phi \simeq \bar\phi+\varphi_s$.
To check this behavior, we now solve numerically the nonlinear Klein-Gordon
equation (\ref{eq:KG-3}).
In practice, we use the physical coordinate $r$, as we focus on the nonlinear
scales that are roughly constant in physical space, as seen in Eq.(\ref{eq:r-sc-def}).
We again subtract the background $\bar\phi$ and compute the nonlinear deviation
$\varphi=\phi-\bar\phi$.

We show the radial profiles at $\tau=1$ of the scalar field, of its spatial and time
derivatives, and of the kinetic argument $\chi$, in Fig.~\ref{fig_dphi}.
We can check that the spatial gradient closely follows the generalized quasistatic
approximation (they cannot be distinguished in the figure).
This also implies that the kinetic argument $\chi$ follows the quasistatic prediction.
Then, at large radii the scalar field $\phi$, and its deviation $\varphi$ from the
background $\bar\phi$, follow the linear solution.
At small radii, the screening mechanism decreases the spatial gradient of the
nonlinear solution. This leads to a flattening of the scalar field in the center
of the halo.

In agreement with these behaviors, the time derivative of the deviation
$\varphi$ is much smaller than for the background $\bar\phi$ and
it follows the linear prediction at large radii. At $\tau=1$ it remains very close
to the linear model down to the center of the halo, as shown by the small value
of $\partial_\tau(\varphi-\varphi_L)_r$.
Thus, the value of the scalar field at the center, and its time drift, follow the cosmological
background.

\subsection{Analytical discussion}
\label{sec:analytical}

We can understand the numerical behaviors found in Fig.~\ref{fig_dphi}
from a simple analysis.
As for the linear case, it is convenient to subtract the cosmological background
by defining the nonlinear deviation $\varphi$, which is not necessarily small,
\beq
\phi = \bar\phi + \varphi ,
\label{eq:background-split-NL}
\eeq
Then, the nonlinear Klein-Gordon equation can be written in a form similar
to the quasistatic Poisson equation (\ref{eq:qsK}) as
\beqa
\nabla( K' \nabla \varphi) & = & 12 \beta \frac{1+\delta-K'}{\tau^2}
+ K' \left( \frac{\partial^2\varphi}{\partial\tau^2}
+ \frac{4}{\tau} \frac{\partial\varphi}{\partial\tau} \right) \nonumber \\
&& + (\partial_\tau K') \left( \frac{\partial\varphi}{\partial\tau} - \frac{4\beta}{\tau} \right) .
\label{eq:KG-5}
\eeqa
For $K'=1$ we recover the linear evolution equation (\ref{eq:KG-linear-pert}).
When we neglect the time derivatives we recover the quasistatic Poisson equation
(\ref{eq:qsK}), using $1+\delta-K' \simeq \delta$ in the high-density core.

Because the screening transition $x_{\rm sc}$ appears very far inside the horizon,
and in fact far inside the overdense region $x_s$ and inside the spatial domain $x_{\rm qs}$,
we are far inside the quasistatic regime, where spatial gradients dominate over time derivatives.
This could also be seen in Fig.~\ref{fig_dphi} above.
Therefore, the dominant terms in the nonlinear Klein-Gordon equation
(\ref{eq:KG-5}) are the spatial derivatives that also appeared in the quasistatic Poisson
equation (\ref{eq:qsK}) and to obtain analytical estimates for the time drift of the scalar field
it is convenient to treat other terms as external sources.
Moreover, because the matter density profile is stationary in physical coordinate $r$,
it is useful to switch to the coordinates $\{r,\tau\}$ that are appropriate on astrophysical scales.
This avoids artificial time dependencies due to comoving coordinates.
Then, integrating once Eq.(\ref{eq:KG-5}) over the radius, as in the Poisson equation
(\ref{eq:phis-screening}), we obtain
\beqa
&& K' \frac{\partial\varphi}{\partial r} = \frac{12\beta}{3-\gamma} r
\left( \frac{r}{x_\star} \right)^{-\gamma} + \frac{1}{\tau^4 r^2} \int_0^r dr \, r^2  \nonumber \\
&& \times \left\{ 12\beta \frac{1-K'}{\tau^2} + K' \left( \frac{\partial^2\varphi}{\partial\tau^2}
+ \frac{4}{\tau} \frac{\partial\varphi}{\partial\tau} \right)_x \right. \nonumber \\
&& \left. + (\partial_\tau K')_x \left( \frac{\partial\varphi}{\partial\tau}
- \frac{4\beta}{\tau} \right)_x \right\} ,
\label{eq:KG-6}
\eeqa
where we explicitly integrated the power-law density contrast (\ref{eq:delta-r})
and the terms with the subscript ``x'' are time derivatives at fixed comoving coordinate $x$.
The time drift of the scalar field deviation $\varphi$ arises from the time-dependent
terms on the right-hand side and from the time dependence that is implicitly
included in the factor $K'$ on the left-hand side, through the kinetic argument
$\chi$ that reads from Eq.(\ref{eq:chi-r}) as
\beqa
\chi & = & \frac{1}{2} \left[ \frac{1}{\tau^4} \left( \partial_\tau \varphi - \frac{4\beta}{\tau} \right)^2
+ \frac{4 r}{\tau^5} \left( \partial_\tau \varphi - \frac{4\beta}{\tau} \right) \partial_r \varphi
\right. \nonumber \\
&& \left. + \left( \frac{4 r^2}{\tau^6} -1 \right) ( \partial_r \varphi )^2 \right] .
\label{eq:chi-r1}
\eeqa

On subhorizon scales the nonlinear Klein-Gordon equation (\ref{eq:KG-6})
is within the quasistatic regime and it is dominated by the left-hand side and the first term
on the right-hand side, which converge to the static solution
$K'[-(\partial_r\varphi)^2/2] \partial_r\varphi = 12\beta r (r/x_\star)^{-\gamma}/(3-\gamma)$.
Then, the time drift from the linear solution that holds at large radii can be obtained
by taking the time derivative of Eq.(\ref{eq:KG-6}). This removes the static matter density
profiles, which does not contribute to the time drift of the scalar field, and the linear solution
that is constant on small scales as seen in Fig.~\ref{fig_dphi_L}.

The first contribution, denoted by the subscript ``1'', associated with the time-dependent
terms on the right-hand side, gives the estimate
\beq
r \gtrsim r_{\rm sc} : \;\;\;
\frac{\partial^2\varphi_1}{\partial r \partial\tau} \sim \frac{\beta r_{\rm sc}^3}{\tau^7 r^2} .
\eeq
Here we used $| \varphi | \ll \beta$ on relevant scales, so that the bracket in
the right-hand side in Eq.(\ref{eq:KG-6}) is dominated by the first and third terms,
which vanish in the linear case $K'=1$ and at radii greater than the screening transition
$r_{\rm sc}$, and we assumed $(\partial_\tau K')_x \sim K' / \tau$.
As explained in section~\ref{sec:generalized-quasistatic} and Fig.~\ref{fig_chi_s},
the radial and time profiles of $K'$ remain smooth even if $K'(\chi)$ is a very steep
or discontinuous function of $\chi$. This regularizes  the last term in Eq.(\ref{eq:KG-6})
We checked that the numerical computations satisfy these properties.
This gives
\beq
\left. \partial_\tau \varphi_1 \right|_{r_{\rm sc}} \sim \frac{\beta r_{\rm sc}^2}{\tau^7}
\label{eq:phi1}
\eeq
and hence
\beq
\left| \partial_\tau \varphi_1 \right|_{r_{\rm sc}} \ll \left| \frac{d\bar\phi}{d\tau} \right| \;\;\;
\mbox{for} \;\;\; \tau \sim 1, \;\; r_{\rm sc} \ll 1 .
\eeq
For the numerical value (\ref{eq:r-sc-num}), this yields at $\tau = 1$ the estimate
$\partial_\tau \varphi_1 \sim 10^{-16}$. This corresponds to the large-radius
tail at $r \gtrsim 10^{-7}$ for $\partial_\tau\varphi - \partial_\tau\varphi_L$
in the lower right panel in Fig.~\ref {fig_dphi}.

The second contribution, denoted by the subscript ``2'', is associated with the time-dependent
factor $K'$ on the left-hand side.
Indeed, the time dependence in the kinetic argument (\ref{eq:chi-r1}) implies that the
radius of the screening transition slightly moves with time and the scalar field must adjust
to this motion.
This can be estimated by writing
\beq
\frac{\partial}{\partial\tau} \left[ K' \frac{\partial \varphi}{\partial r} \right] \simeq 0 ,
\label{eq:quasistatic-time}
\eeq
which expresses that the scalar field follows the quasistatic equilibrium set by the balance
of the left-hand side with the first term on the right-hand side of Eq.(\ref{eq:KG-6}).
Around the screening transition, we can approximate $\chi$ in Eq.(\ref{eq:chi-r1}) by
\beq
\chi \simeq \frac{8\beta^2}{\tau^6} - \frac{1}{2} (\partial_r\varphi)^2
\simeq - \frac{1}{2} (\partial_r\varphi)^2 \simeq \chi_{\rm sc} .
\label{eq:chi-sc}
\eeq
In the first expression we kept both the leading gradient term and the first subleading
correction, which depends on time through the cosmological background.
Then, Eq.(\ref{eq:quasistatic-time}) yields
\beq
K'' \left( - \frac{48\beta^2}{\tau^7} - \frac{\partial\varphi}{\partial r}
\frac{\partial^2\varphi}{\partial r\partial\tau} \right) \frac{\partial \varphi}{\partial r}
+ K' \frac{\partial^2 \varphi}{\partial r\partial\tau} \simeq 0 .
\eeq
At the transition, we have $| K'' | \sim | \Delta K'/\Delta \chi | \gg | K' / \chi |$
as we consider a sharp transition with $|\Delta\chi | \ll | \chi_{\rm sc} |$.
Then, using the last relation (\ref{eq:chi-sc}),
we obtain
\beq
\frac{\partial^2 \varphi}{\partial r\partial\tau} \simeq - \frac{48 \beta^2}{\tau^7 \sqrt{-2\chi_{\rm sc}}} .
\eeq
This yields for this second contribution $\varphi_2$
\beq
\left. \partial_\tau \varphi_2 \right|_{r_{\rm sc}}
\sim \frac{48 \beta^2 r_{\rm sc}}{\tau^7 \sqrt{-2\chi_{\rm sc}}}
\label{eq:phi2}
\eeq
and hence
\beq
\left| \partial_\tau \varphi_2 \right|_{r_{\rm sc}} \ll \left| \frac{d\bar\phi}{d\tau} \right| \;\;\;
\mbox{for} \;\;\; \tau \sim 1, \;\; r_{\rm sc} \ll 1 , \;\; | \chi_{\rm sc} | \gg 1 .
\eeq
For the numerical values (\ref{eq:chis-num}), this yields at $\tau=1$ the estimate
$\partial_\tau\varphi_2 \sim  3 \times 10^{-11}$.
This corresponds to the steep growth of $\partial_\tau\varphi - \partial_\tau\varphi_L$
in the lower right panel in Fig.~\ref{fig_dphi} at $r_{\rm sc}$.
Indeed, this contribution arises at the transition, due to the motion of the screening boundary.

We can note that the two contributions (\ref{eq:phi1}) and (\ref{eq:phi2}), and the
linear solution (\ref{eq:time-derivative-r}), all have different scalings, as they arise from
different terms and physical effects.
However, they all remain much below the background time derivative at late times.
This confirms that the scalar field remains strongly coupled to the cosmological
background for small-scale matter overdensities.

Thus, we find that the naive local analysis of the equation of motion (\ref{eq:KG-1}),
which could suggest that in screened regions where $K'$ is very large
the scalar field $\phi$ no longer evolves and remains constant in space and time,
is not correct.
In fact, the only size that can be considered local is the Hubble radius,
independently of the variations and nonlinearities of $K'$.
This could be expected from the fact that the propagation speed remains
of order unity, even in nonlinear domains, and that there is no damping of the
amplitude of the scalar field as the equation of motion only involves
its derivatives. Indeed, in small-scale nonlinear environments the radial propagation
speed reads  \cite{Brax:2014c}
\beq
c_{\phi}^2 = \frac{K'+2\chi K''}{K'}  \geq 1 .
\eeq
It is always greater than unity for the models that we consider here, because $\chi \leq 0$
in the spatial domain and $K'' \leq 0$ as $K'$ shows a monotonic decrease from
$K'_{\rm sc}$ down to unity.
More generally, it is typically of order unity, as for power-law kinetic functions we have
$\chi K'' \sim K'$ whereas $\chi K'' \ll K'$ in regimes where $K'$ is almost constant.
However, in the middle of the nonlinear transition for a sharp kinetic function $K'$,
we can have $\chi K'' \gg K'$ and $c_\phi \gg 1$.
In any case, the lower bound $c_\phi \geq 1$ implies that the quasistatic approximation
applies on all subhorizon scales, including the nonlinear regime.
Then, the gradient of the scalar field is set by the nonlinear Poisson equation,
which follows from the quasistatic approximation of the nonlinear Klein-Gordon equation,
while the cosmological background sets the boundary condition at the horizon.
This makes the cosmological time drift apply on all scales, down to the center of the halos.

The validity of the quasistatic approximation also prevents the long-term development
of fast-moving caustics, despite the spatially varying sound speed. In fact, we have seen
that the nonlinear Poisson equation automatically smooths radial profiles and
discontinuities of $K'(\chi)$ do not lead to discontinuous scalar profiles
and radial gradients (only the second-order radial derivative would be discontinuous).
The nonlinear transition can lead to steep increases for the first-order time derivative
at the transition, but their magnitude remains very small and much below the time derivative
of the cosmological background. In particular, the kinetic argument $\chi$ remains
in the spatial domain, $\chi < 0$.

One might try to circumvent the coupling to the cosmological background
with a model such that $c_\phi \simeq 0$ over an intermediate range of $\chi$.
This could invalidate the quasistatic approximation and separate the inner and outer
domains.
However, $c_\phi^2 < 0$ leads to gradient instabilities so that well-behaved models
typically have $c_\phi^2 > 0$.
Then, one may consider models where $c_\phi^2$ remains positive but becomes
sufficiently small over some range to invalidate the quasistatic approximation.
However, this involves a fine-tuning, as it requires $K' \propto 1/\sqrt{-\chi}$
over this range, which uniquely sets the kinetic function up to a proportionality factor
and subleading corrections.
We do not investigate this case further in this paper.

\section{Conclusion}
\label{sec:Conclusion}

The value of the scalar field deep inside a collapsed region of the Universe is highly
relevant as it determines the value of Newton's constant, which is proportional to
$A^2(\phi)$ where $A(\phi)\sim e^{\beta \phi/M_{\rm Pl}}$ is the coupling function to matter
and $\beta={\cal O}(1)$ the coupling to matter. In screened regions where the K-mouflage
mechanism is at play, the spatial gradients of the scalar field are large, much larger than
the time derivatives, and the fifth force induced by the scalar is largely depleted.
On the other hand, it is well known that a linear time drift $H_0 t$ still allows for
static solutions around a time-independent astrophysical object and can provide an
approximate matching with the large-scale cosmological evolution of the scalar field.
This induces then a cosmological time drift of Newton's constant, jeopardizing the viability
of many models of the K-mouflage type.

In this paper, we have investigated the influence of the background cosmology
on the short-distance physics within a collapsed structure of the Universe.
We have taken it to be described by a self-similar power-law density profile,
which allows us to provide an almost exact treatment.
We find that inside the structure there is a critical radius
$x_{\rm qs}$ within which the quasistatic approximation holds, in the sense that
spatial derivatives are greater than time derivatives.
This radius is much smaller than the size $x_s$ of the matter overdensity, where
the matter density contrast becomes of order unity.
However, spatial gradients are well described by the quasistatic approximation
up to the horizon, and hence up to much larger scales, as found for other
modified-gravity scenarios in previous studies.
We also find that for structures that grow fast with time, which could apply
to transient mergings but also to the fast building of the cosmic web at
redshifts $z \gtrsim 2$, the time derivative of the scalar field perturbations
remains greater than its spatial gradient.

Screening of the fifth force takes place only well inside the quasistatic radius,
where $\nabla \phi \gg \partial_\tau \phi$.
However, inside the screening radius $x_{\rm sc}$ and down to the center of the
overdensity, the values of the scalar field remain strongly dependent on the background
cosmological evolution: no screening of the time drift of Newton's constant takes place.
The scalar field only decouples from the cosmological
background if the matter structure extends up to the horizon, which is not the case
for realistic astrophysical and cosmological structures.
Of course, this result does not invalidate
K-mouflage models and simply implies that the strong constraints deduced in
\cite{Barreira2015} must be taken seriously.
Thus, the K-mouflage screening mechanism only damps the spatial gradients of the
scalar field, reducing the fifth force in small-scale high-density environments,
while following the large-scale drift of the cosmological background.
We can expect that this behavior extends to other derivative screening mechanisms,
such as Vainshtein screening.

Thus, we have shown that the dynamics of screening in K-mouflage models
is more complex than can be deduced by a fully quasistatic approximation.
In particular, the appearance of two radii: the quasistatic and screening radii
is a new feature. It would be extremely interesting to see if N-body simulations
of K-mouflage models could reveal other new dynamical characteristics of K-mouflage,
for instance around fast-growing structures.
This is left for future work.

\bibliography{ref1}   

\end{document}